\documentclass[11pt,preprint]{aastex}

\def\dd{{\rm d}}

\def\be{\begin{equation}}
\def\ee{\end{equation}}

\def\sT{\sigma_{\rm T}}

\def\tauCoul{\tau_{\rm Coul}}
\def\taulife{\tau_{\rm life}}
\def\Ya{Y_\alpha}
\def\dM{\dot{M}}
\def\dMeq{\dot{M}_{\rm eq}}
\def\Tnuc{T_*}
\def\Rnuc{R_*}
\def\tnuc{\tau_*}
\def\treac{\tau_{\rm reac}}
\def\taub{\tau_\beta}
\def\dY{\dot{Y}}

\def\tbet{\tilde{\beta}}
\def\taucoll{\tau_{\rm coll}}
\def\tcoll{t_{\rm coll}}
\def\tc{t_+}
\def\tbsp{\tilde{\beta}_{\rm sp}}
\def\tlife{t_{\rm life}}

\def\Gsp{\Gamma_{\rm sp}}
\def\Gdec{\Gamma_{\rm dec}}
\def\OmK{\Omega_{\rm K}}

\def\Td{T_{\rm deg}}
\def\wdeg{w_{\rm deg}}
\def\Gsat{\Gamma_{\rm max}}
\def\thd{\theta_{\rm deg}}
\def\Tmax{T_{\rm max}}
\def\Tn{T_n}
\def\tmix{t_{\rm mix}}
\def\dn{\dot{n}}
\def\lbar{\lambda\llap {--}}
\def\om{\omega}
\def\Lth{L_{\rm th}}
\def\LP{L_{\rm P}}
\def\etath{\eta_{\rm th}}
\def\etaP{\eta_{\rm P}}
\def\ssp{\sigma_{\rm sp}}
\def\phisp{\phi_{\rm sp}}
\def\phidec{\phi_{\rm dec}}
\def\tgg{\tau_{\gamma\gamma}}
\def\rin{r_{\rm in}}
\def\wmax{w_{\rm max}}
\def\fmax{f_{\rm max}}

\newbox\grsign \setbox\grsign=\hbox{$>$} \newdimen\grdimen \grdimen=\ht\grsign
\newbox\simlessbox \newbox\simgreatbox \newbox\simpropbox
\setbox\simgreatbox=\hbox{\raise.5ex\hbox{$>$}\llap
     {\lower.5ex\hbox{$\sim$}}}\ht1=\grdimen\dp1=0pt
\setbox\simlessbox=\hbox{\raise.5ex\hbox{$<$}\llap
     {\lower.5ex\hbox{$\sim$}}}\ht2=\grdimen\dp2=0pt
\setbox\simpropbox=\hbox{\raise.5ex\hbox{$\propto$}\llap
     {\lower.5ex\hbox{$\sim$}}}\ht2=\grdimen\dp2=0pt
\def\simgt{\mathrel{\copy\simgreatbox}}
\def\simlt{\mathrel{\copy\simlessbox}}

\begin{document}

\title{Nuclear composition of gamma-ray burst fireballs}

\author{Andrei M. Beloborodov\altaffilmark{1,2,3}}

\altaffiltext{1}{Canadian Institute for Theoretical Astrophysics,
60 St. George Street, Toronto, ON M5S 3H8, Canada}

\altaffiltext{2}{Physics Department, Columbia University, 538  West 120th
Street New York, NY 10027}

\altaffiltext{3}{Astro-Space Center of Lebedev Physical
Institute, Profsojuznaja 84/32, Moscow 117810, Russia}

\begin{abstract}
We study three processes that shape the nuclear composition of GRB
fireballs: (1) neutronization in the central engine,
(2) nucleosynthesis in the fireball as it expands and cools, and
(3) spallation of nuclei in subsequent internal shocks. The fireballs are
found to have a neutron excess and a marginally successful nucleosynthesis.
They are composed of free nucleons, $\alpha$-particles, and deuterium.
A robust result is the survival of a significant neutron component,
which has important implications. First, as shown in previous works,
neutrons can lead to observable multi-GeV neutrino emission.
Second, as we show in an accompanying paper, neutrons impact
the explosion dynamics at radii up to $10^{17}$~cm and
change the mechanism of the GRB afterglow emission.
\end{abstract}



\section{Introduction}

Cosmological gamma-ray bursts (GRBs) are powerful explosions in distant 
galaxies. A physical picture of these phenomena has emerged in the past 
decade (see M\'esz\'aros 2002 for a review): the GRBs are generated by 
compact, dense, and energetic engines, and they are likely related to black 
hole formations. The typical mass of the central engine is believed to be 
a few $M_\odot$, its size is $10^6-10^7$~cm, and its temperature is
$kT=1-10$~MeV. The engine ejects a hot outflow (``fireball'') 
made of free nucleons, $e^\pm$ pairs, trapped blackbody radiation, and 
magnetic fields.

The initial nuclear composition
of the fireball and its evolution during expansion turn out crucial for the
explosion physics. In particular, neutrons were shown to be an important
component in the explosion (Derishev, Kocharovsky, \& Kocharovsky 1999)
and an extreme case of fireballs with a large neutron excess was studied
(Fuller, Pruet, \& Abazajian 2000). Particular attention was given to the 
relative motion of the neutron and ion components of the fireball, which can 
lead to observable multi-GeV neutrino emission (Bahcall \& M\'esz\'aros 2000; 
M\'esz\'aros \& Rees 2000). The two components can be ejected with 
substantially different velocities, and this can affect the observed 
afterglow emission of GRBs (Pruet \& Dalal 2002; Bulik, Sikora, \& Moderski 
2002). In an accompanying paper (Beloborodov 2002, hereafter Paper~2) we 
show that the presence of neutrons crucially changes the fireball interaction 
with an external medium and implies a new mechanism of the afterglow 
production.

The present paper focuses on processes that shape the nuclear composition of 
GRB fireballs. In \S~2 we study the production of neutrons by the central 
engine and the resulting neutron-to-proton ratio in the ejected material
(scale $R\sim r_0\simlt 10^{7}$~cm). In \S~3, we calculate nuclear reactions 
in the expanding fireball and find abundances of survived free nucleons and 
synthesized helium ($R=10^7-10^9$~cm). Similar nucleosynthesis calculations, 
with different codes, have been done recently by Lemoine (2002) and Pruet, 
Guiles, \& Fuller (2002). In \S~4, we study spallation of helium at later 
stages when internal motions develop in the fireball ($R=10^{9}-10^{12}$~cm).
The subsequent dynamics of GRB blast waves with the survived neutron
component ($R=10^{15}-10^{17}$~cm) is investigated in Paper~2.


\section{Neutronization}

There are various models for the central engines of GRBs.
It may be a young neutron star whose rotational energy is emitted
in a magnetized wind, a neutron star merger, or a massive star collapse.
The latter two scenarios proceed via the formation of a black hole of
mass $M\sim M_\odot$ and subsequent disk-like accretion of a comparable mass.
The baryonic component of the ejected fireball is then picked up from
the accretion disk.

The central engines are sufficiently dense for the electron capture reaction.
We will calculate at what densities and temperatures this process creates
a neutron excess (neutron-to-proton ratio above unity), and then show
that GRB engines are likely to satisfy these conditions.
We will illustrate with the accretion-type models of GRBs,
where the matter density can be relatively low and neutronization is most
questionable.

\subsection{The equilibrium $Y_e$}

Consider a dense, $\rho>10^{7}$~g~cm$^{-3}$, and hot, $kT>m_ec^2$, matter.
The rates of photon emission and absorption are huge and the matter
is filled with Planckian radiation. Also, the rates of $e^\pm$ pair
creation and annihilation ($\gamma+\gamma\leftrightarrow e^-+e^+$) are huge,
and the pairs are in perfect thermodynamic equilibrium with the baryonic
matter and radiation. The $e^\pm$ number densities are
\be
\label{eq:npm}
  n_\pm=\frac{(m_ec)^3}{\pi^2\hbar^3}\int_0^\infty
           f_\pm(\sqrt{p^2+1}){p^2\dd p}.
\ee
Here $p$ is particle momentum in units of $m_ec$ and $f_\pm$ is the
Fermi-Dirac occupation function,
\be
\label{eq:fpm}
  f_\pm(x)=\frac{1}{\exp[(x-\mu_\pm)/\theta]+1}
          =\frac{1}{\exp[(x\pm\mu)/\theta]+1},
\ee
where $\theta=kT/m_ec^2$ and $\mu\equiv\mu_-=-\mu_+$ is the electron chemical
potential in units of $m_ec^2$ (thermodynamic equilibrium of $e^\pm$ with
radiation implies $\mu_++\mu_-=0$). In addition to equation~(\ref{eq:npm}),
we have the charge-neutrality condition,
\be
\label{eq:charge}
  n_--n_+=Y_e\frac{\rho}{m_p},
\ee
where $Y_e$ is the proton-to-nucleon ratio, which would equal the
electron-to-nucleon ratio in the absence of $e^\pm$ pairs.
Equations~(\ref{eq:npm}) and (\ref{eq:charge}) determine $\mu$ and $n_\pm$ 
for given $T$, $\rho$, and $Y_e$. The electrons become degenerate if $\mu$ 
exceeds $\theta$, which happens below the characteristic degeneracy 
temperature,
\be
\label{eq:Td}
 \thd=\frac{\hbar}{m_ec}\left(\frac{\rho}{m_p}\right)^{1/3}, \qquad
  k\Td=7.7\left(\frac{\rho}{10^{11}{\rm g~cm}^{-3}}\right)^{1/3} {\rm ~MeV}.
\ee
Degeneracy exponentially suppresses the positron density,
$n_+/n_-\approx \exp(-\mu/\theta)$, because then $e^\pm$ are created
only at energies $E/m_ec^2\sim \mu>\theta$, in the exponential
tail of the thermal distribution.

At temperatures and densities under consideration, the baryonic matter is 
in nuclear statistical equilibrium, and it is dominated by free nucleons 
in the unshadowed region of Figure~1. The boundary of this region has been 
calculated with the Lattimer-Swesty Equation of State code (Lattimer \& 
Swesty 1991).  The free protons and neutrons can capture $e^-$ and $e^+$ via 
charged current reactions,
\begin{equation}
  e^-+p\rightarrow n+\nu, \qquad e^++n\rightarrow p+\bar{\nu}.
\end{equation}
These reactions can rapidly convert protons into neutrons and neutrons
back into protons, and establish an equilibrium $Y_e=n_p/(n_n+n_p)$,
where $n_p$ and $n_n$ are number densities of protons and neutrons,
respectively.\footnote{The neutron decay $n\rightarrow p+e^-+\bar{\nu}$
is a slow process on GRB timescales and it is neglected here.}
We now calculate the equilibrium $Y_e(T,\rho)$, and
in \S~2.2 we will show how it applies to GRB central engines.

The exact equilibrium $Y_e$ depends on whether the opposite reactions ---
reabsorption of the emitted $\nu$ and $\bar{\nu}$ --- are also significant.
We first consider the $\nu$-transparent case, where reabsorption can be
neglected, and then address the $\nu$-opaque case.

\subsubsection{Neutrino-transparent matter}

The rates of $e^-$ and $e^+$ capture can be derived from the
standard electro-weak theory (e.g., Shapiro \& Teukolsky 1983, Bruenn 1985). 
We assume the nucleons to be non-degenerate, and then the rates are
\be
\label{eq:e-p}
  \dn_{e^-p}=Kn_p\int_0^\infty f_-(\om+Q)(\om+Q)^2
 \left[1-\frac{1}{(\om+Q)^2}\right]^{1/2}\om^2\dd\om,
\ee
\be
\label{eq:e+n}
  \dn_{e^+n}=Kn_n\int_{Q+1}^\infty f_+(\om-Q)(\om-Q)^2
      \left[1-\frac{1}{(\om-Q)^2}\right]^{1/2}\om^2\dd\om,
\ee
where $\om$ is neutrino energy in units of $m_ec^2$, $Q=(m_n-m_p)/m_e=2.531$,
and $K\approx 6.5\times 10^{-4}$~s$^{-1}$. The constant $K$ can be expressed 
in terms of the mean lifetime of neutrons with respect to $\beta$-decay, 
$\tau_\beta\approx 900$~s, as $K\approx (1.7\tau_\beta)^{-1}$ 
(Shapiro \& Teukolsky 1983, pp.~316, 524).

An equilibrium $Y_e$ is established when the rates of $e^-$ and $e^+$ 
captures are equal,\footnote{In the $\nu$-transparent regime, neutrinos 
are sometimes prescribed a zero chemical potential, and the balance
$\mu+\mu_p=\mu_n+\mu_\nu$ is used with $\mu_\nu=0$ to determine the 
equilibrium $Y_e$.  In fact, the balance of chemical potentials
does not hold because the neutrinos are out of thermodynamic equilibrium.
This balance would be valid only in the cold limit $T\rightarrow 0$
(Landau \& Lifshitz 1980).}
\be
\label{eq:balance}
  \dn_{e^+n}=\dn_{e^-p}.
\ee
Equations~(\ref{eq:npm}), (\ref{eq:charge}), and (\ref{eq:balance}) determine
$Y_e$ for given $\rho$ and $T$. Contours of the function $Y_e(T,\rho)$ on
the $T-\rho$ plane are shown in Figure~1.

\begin{figure}
\plotone{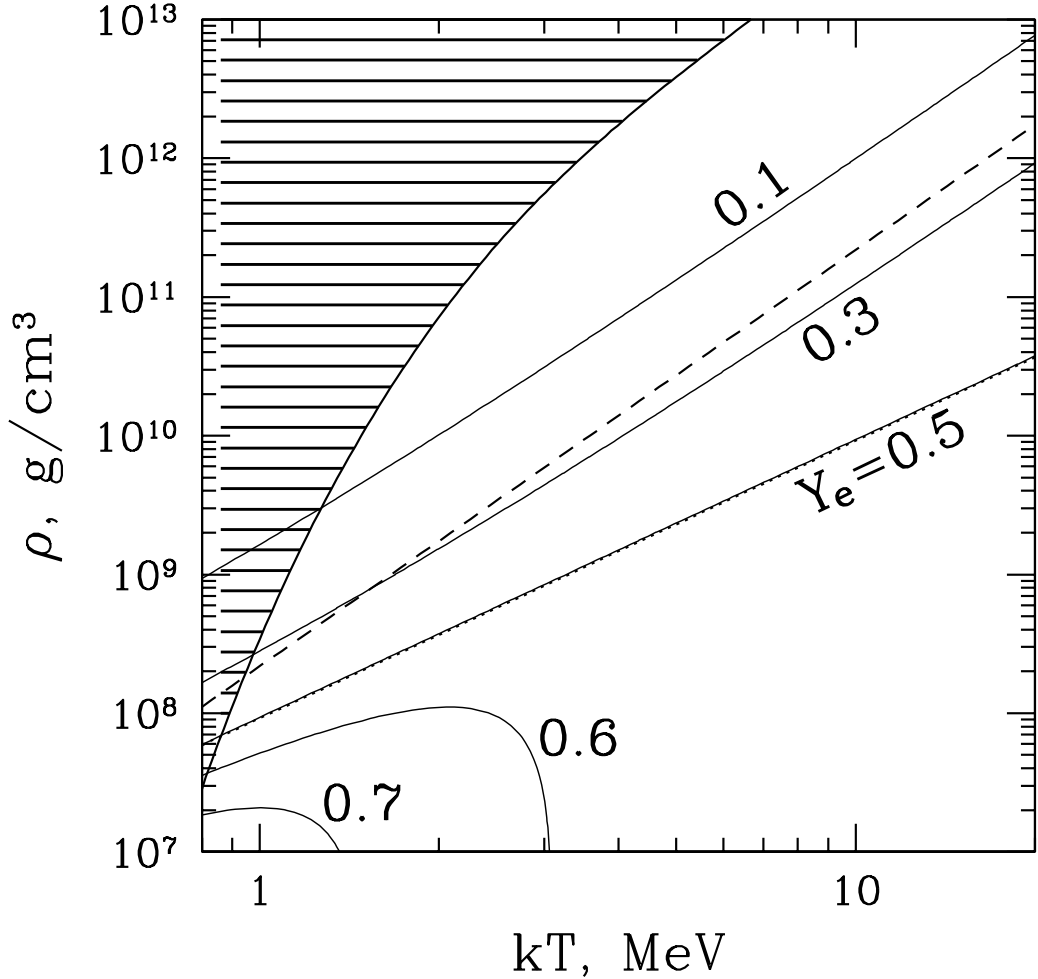}
\caption{Contours of the equilibrium $Y_e(T,\rho)$ on the $T-\rho$ plane
for $\nu$-transparent matter. Dashed line shows the degeneracy temperature
(eq.~\ref{eq:Td}). In the shadowed region, the baryonic matter is dominated
by composite nuclei, and the calculations based on rates~(\ref{eq:e-p}) and
(\ref{eq:e+n}) are not valid. The n/p-ratio equals $(1-Y_e)/Y_e$,
and the neutrons dominate over protons where $Y_e<0.5$.
The analytically calculated  boundary $Y_e=0.5$
(eq.~\ref{eq:Tn_tr}) is also plotted here by a dotted line, which perfectly
coincides with the numerically found contour $Y_e=0.5$.
\label{fig1}}
\end{figure}

The transition from a proton excess ($Y_e>0.5$) to a neutron excess 
($Y_e<0.5$) takes place in the region of mild degeneracy where $\mu<\theta$. 
We now focus on this region and derive the equilibrium $Y_e$ analytically. 
At $\mu<\theta$ and $\theta>Q+1$ equations~(\ref{eq:e-p}) and (\ref{eq:e+n}) 
simplify,
\be
\label{eq:e-p_}
 \dn_{e^-p}=Kn_p\theta^5\left[\frac{45}{2}\zeta(5) +
 \frac{7\pi^4}{60}\frac{(2\mu-Q)}{\theta}\right],
\ee
\be
\label{eq:e+n_}
  \dn_{e^+n}=Kn_n\theta^5\left[\frac{45}{2}\zeta(5) -
 \frac{7\pi^4}{60}\frac{(2\mu-Q)}{\theta}\right],
\ee
where $\zeta(5)=1.037$ is Riemann $\zeta$-function. We neglected here 
next-order terms $O(Q^2/\theta^2)$, $O(\mu^2/\theta^2)$, and 
$O[(Q+1)^5/\theta^5]$, and used the formula
$\int_0^\infty (e^x+1)^{-1}x^n\dd x=(1-2^{-n})\Gamma(n+1)\zeta(n+1)$
with $\Gamma(n+1)=n!$ for integer $n$.
Equating the two rates, we get the equilibrium $Y_e=n_p/(n_n+n_p)$,
\be
\label{eq:Ye_tr}
   Y_e=\frac{1}{2}+\frac{7\pi^4}{1350\zeta(5)}\frac{(Q/2-\mu)}{\theta}
      =\frac{1}{2}+0.487\frac{(Q/2-\mu)}{\theta}.
\ee
In the non-degenerate limit, $\mu/\theta\rightarrow 0$, this gives
$Y_e=0.5+0.616/\theta>0.5$ and implies a proton excess, which is due to the
positive difference $Q$ between the neutron and proton masses. A very mild 
degeneracy $\mu=Q/2<\theta$ is sufficient to drive $Y_e$ below 0.5. This 
happens because the $e^+$ density is reduced by the degeneracy effects and
the $e^-$ capture becomes preferential. 

It is instructive to write the $e^-$ and $e^+$ densities using the linear 
expansion of equation~(\ref{eq:npm}) in $\mu/\theta$,
\be
\label{eq:expan}
  n_\pm=\frac{1}{\pi^2\lbar^3}\left[\frac{3}{2}\zeta(3)\theta^3\
     \mp\frac{\pi^2}{6}\mu\theta^2\right], \qquad \mu<\theta,
\ee
where $\lbar=\hbar/m_ec=3.862\times 10^{-11}$~cm and $\zeta(3)=1.202$.
Then equation~(\ref{eq:charge}) yields
\be
\label{eq:mu_e}
 \mu=3Y_e\frac{\lbar^3\rho}{m_p\theta^2}.
\ee
The condition $\mu>Q/2$ that defines the neutron-excess region on the
$T-\rho$ plane ($Y_e<0.5$) can now be written as $\theta<\theta_n(\rho)$,
where
\be
\label{eq:Tn_tr}
  \theta_n=\left(\frac{3\lbar^3\rho}{Qm_p}\right)^{1/2}, \qquad
  k\Tn=33\rho_{11}^{1/2}{\rm ~MeV}.
\ee
This simple formula perfectly coincides with the numerical results (Fig.~1).
That a very mild electron degeneracy is sufficient to create a neutron 
excess can also be seen by comparing $T_n$ with $\Td$,
\be
  \frac{\Tn}{\Td}\approx 4.3\rho_{11}^{1/6}.
\ee

A useful explicit formula for the equilibrium $Y_e(T,\rho)$ in 
$\nu$-transparent matter with mild degeneracy is derived from 
equations~(\ref{eq:Ye_tr}) and (\ref{eq:mu_e}),
\be
\label{eq:Ye_opaque}
  Y_e(\theta,\rho)=\frac{1}{2}\left(1+\frac{0.487Q}{\theta}\right)
    \left(1+1.46\frac{\lbar^3\rho}{m_p\theta^3}\right)^{-1}.
\ee
It agrees with the numerical calculations shown in Figure~1 with
a high accuracy, $\delta Y_e/Y_e<1$\% at $Y_e>0.35$. In a more degenerate
region, where the equilibrium $Y_e<0.35$, the formula can still be used,
though its error increases to 10\% at $Y_e=0.2$ and 30\% at $Y_e=0.1$.

\subsubsection{Neutrino-opaque matter}

If the matter is opaque to the emitted neutrinos, a complete thermodynamic
equilibrium is established. A detailed balance now holds,
$e^-+p\leftrightarrow n+\nu$ and $e^++n\leftrightarrow p+\bar{\nu}$,
and the equilibrium $Y_e$ is determined by the condition
\be
\label{eq:mubal}
   \mu+\mu_p=\mu_n+\mu_\nu,
\ee
where $\mu$, $\mu_p$, $\mu_n$, and $\mu_\nu$ are chemical potentials
(in units of $m_ec^2$) of
the electrons, protons, neutrons, and neutrinos, respectively, all including
the particle rest-mass energy. The antineutrinos
have chemical potential $\mu_{\bar{\nu}}=-\mu_\nu$, so that
$\mu_++\mu_n=\mu_p+\mu_{\bar{\nu}}$ is also satisfied.
The neutrons and protons have Maxwellian distributions, which gives
$n_nn_p^{-1}=\exp[(\mu_nm_ec^2-m_nc^2)/kT]\exp[(-\mu_pm_ec^2+m_pc^2)/kT]$ and
\be
\label{eq:Maxw}
  \mu_n-\mu_p=\theta\ln(n_n/n_p)+Q.
\ee

The thermalized $\nu$ and $\bar{\nu}$ obey Fermi-Dirac statistics, and 
they are described in the same way as $e^\pm$ (see eqs.~[\ref{eq:npm}] and 
[\ref{eq:fpm}]); the only difference is that the statistical weight of 
energy states is one for neutrinos and two for electrons. The neutrino 
chemical potential vanishes if $\nu$ and $\bar{\nu}$ have equal densities, 
$n_\nu=n_{\bar\nu}$. If, however, the matter emits non-equal numbers of 
$\nu$ and $\bar{\nu}$ (its $Y_e$ is changing) then $n_\nu\neq n_{\bar{\nu}}$ 
and $\mu_\nu\neq 0$.

Suppose $N_\nu$ neutrinos and $N_{\bar{\nu}}$
anti-neutrinos have been emitted per nucleon. This causes a change of $Y_e$,
\be
  Y_e-Y_e^0= N_{\bar{\nu}}- N_\nu,
\ee
where $Y_e^0$ is an initial value that the matter had before the neutrino
emission.  If all the emitted neutrinos remain trapped in the matter then
$(n_\nu-n_{\bar{\nu}})/n_b=Y_e-Y_e^0$, where $n_b=n_n+n_p$ is the total
nucleon density. If a fraction $x$ of the emitted neutrino diffused out of
the matter then $(n_\nu-n_{\bar{\nu}})/n_b=(1-x)(Y_e-Y_e^0)$ and
\be
\label{eq:x}
 \frac{n_\nu-n_{\bar{\nu}}}{n_--n_+}=(1-x)\left(1-\frac{Y_e^0}{Y_e}\right).
\ee
Let us first consider the case $x\rightarrow 1$ (efficient neutrino cooling).
Then $|n_\nu-n_{\bar{\nu}}|\ll n_--n_+$ and the neutrino chemical potential
can be neglected compared to that of the electrons. Thus, the chemical
equilibrium in $\nu$-cooled, $\nu$-opaque matter reads
\be
\label{eq:mubal_}
  \mu=\theta\log(n_n/n_p)+Q.
\ee
Equation~({\ref{eq:mubal_}) combined with equations~(\ref{eq:npm}) and
(\ref{eq:charge}) determines an equilibrium $Y_e$ as a function of $T$ and 
$\rho$. Note however that equation~(\ref{eq:mubal_}) assumes free nucleons, 
which is invalid in the shadowed region of the $T-\rho$ plane shown in 
Figure~1 (the boundary of this region is also plotted in Fig.~2). The 
chemical balance can be extended to this region if $\mu_n$ and $\mu_p$ are 
corrected for the heavy nuclei formation, which we do using the 
Lattimer-Swesty Equation of State code. Then we find the equilibrium $Y_e$ 
that satisfies $\mu=\mu_n-\mu_p$. The results are shown in Figure~2.

\begin{figure}
\plotone{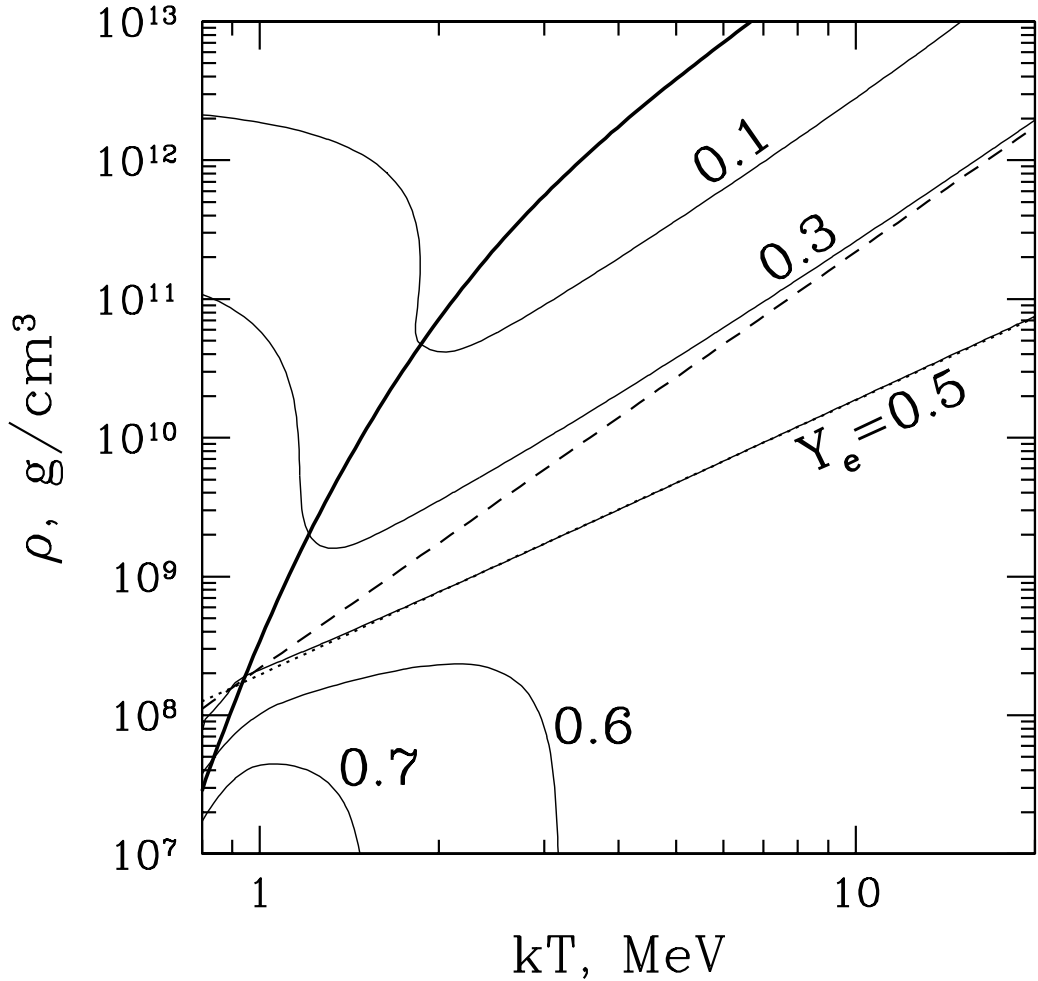}
\caption{Contours of the equilibrium $Y_e(T,\rho)$ on the $T-\rho$ plane
for $\nu$-opaque matter with $\mu_\nu=0$. Dashed line shows the degeneracy
temperature (eq.~\ref{eq:Td}). Thick solid curve is
the boundary of the free-nucleon region (same as in Fig.~1). Above this
curve, the composite nuclei dominate, which prefer equal numbers of
neutrons and protons, and therefore the contours $Y_e={\rm const}$ bend
upward. The analytically calculated  boundary $Y_e=0.5$ (eq.~\ref{eq:Tn_op})
is shown by dotted line; it coincides with the numerically found contour
$Y_e=0.5$.
}
\end{figure}

The neutron excess boundary, $Y_e=0.5$, lies in the region where
all nucleons are free and the electrons are mildly degenerate, $\mu<\theta$.
Equation~(\ref{eq:mubal_}) applies here and shows that this boundary is 
defined by $\mu=Q$. Using the linear expansion in $\mu/\theta$ 
(eq.~\ref{eq:mu_e}), we get a simple equation for $Y_e(\theta,\rho)$,
\be
   \frac{3Y_e}{\theta^2}\frac{\rho\lbar^3}{m_p}
   =\theta\log\left(\frac{1-Y_e}{Y_e}\right)+Q.
\ee
In particular, one sees that $Y_e=0.5$ corresponds to
\be
\label{eq:Tn_op}
   \theta_n=\left(\frac{3\lbar^3\rho}{2Qm_p}\right)^{1/2}, \qquad
   kT_n=23.1\rho_{11}^{1/2}{\rm ~MeV}.
\ee
It coincides exactly with the contour $Y_e=0.5$ calculated with the
Lattimer-Swesty code (Fig.~2). Note also that $T_n=3\Td\rho_{11}^{1/6}$.

Finally, we address the regime where the neutrinos are not only thermalized
but also trapped in the matter ($x\approx 0$ in eq.~[\ref{eq:x}]).
It can happen in GRB accretion flows with high accretion rates, 
$\dM>1M_\odot$/s (e.g. Di Matteo, Perna, \& Narayan 2002).
Then $\mu_\nu$ should not be neglected in the chemical balance.
Like equation~(\ref{eq:mu_e}) we derive for neutrinos
\be
  \mu_\nu=\frac{6\lbar^3(n_\nu-\nu_{\bar{\nu}})}{\theta^2},
\ee
and substitute $n_\nu-\nu_{\bar{\nu}}=(Y_e-Y_e^0)(\rho/m_p)$.
The chemical balance $\mu-\mu_\nu=\mu_n-\mu_p$ now reads
\be
   \frac{3(2Y_e^0-Y_e)}{\theta^2}\frac{\rho\lbar^3}{m_p}
   =\theta\log\left(\frac{1-Y_e}{Y_e}\right)+Q,
\ee
and $Y_e=0.5$ corresponds to
\be
\label{eq:Tn_op_}
  \theta_n=\left[\frac{3(2Y_e^0-0.5)\lbar^3\rho}{Qm_p}\right]^{1/2}.
\ee
At $Y_e^0=0.5$ it coincides with equation~(\ref{eq:Tn_op}) as it should:
$Y_e=0.5=Y_e^0$ requires that $\nu$ and $\bar{\nu}$ are emitted in equal
numbers and $\mu_\nu=0$.

\subsubsection{Effects of magnetic fields}

Magnetic fields have been neglected in the above calculations,
which is a valid approximation as long as the field does not affect the
particle distribution functions. Magnetic fields can be generated in
the GRB central engines by dynamo, and their energy is likely a fraction
$\epsilon_B<1$ of the total energy density $w$, so that
\be
  B=\sqrt{8\pi\epsilon_B w}.
\ee
Here $w$ includes the energy of baryons, radiation, and $e^\pm$;
it also includes the neutrino energy if the disk is $\nu$-opaque.
The field has the strongest effect on light charged particles --- $e^\pm$.
It introduces the discrete energy levels (Landau \& Lifshitz 1980),
\be
 \frac{E_j}{m_ec^2}=\left[1+p_z^2+2j\frac{\hbar\Omega_B}{m_ec^2}\right]^{1/2},
 \qquad j=0,1,...,
\ee
where $-\infty<p_z<\infty$ is the component of the electron momentum parallel
to the field and $\Omega_B=eB/m_ec$. The magnetic
field also changes the phase-space factor in equation~(\ref{eq:npm}):
$p^2\dd p$ is replaced by $(\hbar\Omega_B/m_ec^2)(\dd p_z/2)$. Both effects 
are important if $(\hbar\Omega_B/m_ec^2)>p_z^2$. The mean parallel momentum 
of the relativistic $e^\pm$ equals $\sqrt{3}kT/c$, and the condition for the 
field effects to be important reads
\be
\label{eq:B}
   \frac{\hbar\Omega_B}{m_ec^2}>3\theta^2.
\ee
For any plausible $\epsilon_B<1$, the magnetic field is important only where
the electrons are degenerate, and the energy density $w$ is dominated by
either baryons, $w_b=(3/2)kT\rho/m_p$, or degenerate electrons,
$\wdeg=(9/4)(\pi/3)^{4/3}\hbar c(Y_e\rho/m_p)^{4/3}$.
The condition~(\ref{eq:B}) can be written as
\be
\label{eq:B_}
\theta<\left(\frac{\rho\lbar^3}{m_p}\right)^{1/3}\times
\left\{\begin{array}{ll}
(4\pi\alpha_f\epsilon_B/3)^{1/3} & \epsilon_B>(\pi^3Y_e^4/32\alpha_f),\\
(\pi/3)^{7/12}(6\alpha_f\epsilon_B)^{1/4}Y_e^{1/3} &
  \epsilon_B<(\pi^3Y_e^4/32\alpha_f).\\
       \end{array}\right.
\ee
where $\alpha_f$ is the fine structure constant. The upper line corresponds
to $w_b>\wdeg$, and the lower line to $\wdeg>w_b$.

\subsection{GRB central engines}

There exists a general constraint on the electron degeneracy in the engine.
Its derivation makes use of two facts: (1) The engine is gravitationally 
bound --- otherwise it would explode, and a high baryon contamination of 
the fireball would be inevitable. GRB models normally envision a quasi-static 
engine that liberates a fraction of its gravitational binding energy and 
passes it to a tiny amount of mass outside the engine, thus creating a highly 
relativistic outflow. (2) The engine is compact (size $r<10^7$~cm), and it 
has a relativistic blackbody temperature $\theta=kT/m_ec^2>1$.

The sound speed in a gravitationally bound object of size $r$ and mass $M$
must be smaller than $(GM/r)^{1/2}$. It gives the constraint
\be
\label{eq:constr}
  \frac{P}{\rho}\simlt 0.1c^2\left(\frac{r}{3r_g}\right)^{-1},
\ee
where $r_g=2GM/c^2$ is the gravitational radius of the object. Pressure
$P=P_\gamma+P_\pm+P_\nu+P_b$ includes contributions from radiation, $e^\pm$,
neutrinos, and baryons. We have $P_\gamma+P_\pm=(11/12)aT^4$ if the $e^\pm$
are weakly degenerate, and a maximum $P_\nu=(7/24)aT^4$ for each neutrino
species if it is thermalized, where
$a=(\pi^2k^4/15\hbar^3c^3)=7.56\times 10^{-15}$~erg~cm$^{-3}$~K$^{-4}$ is
the radiation constant. Approximately,
\be
\label{eq:P}
   P\approx aT^4+\frac{\rho}{m_p}kT.
\ee
When comparing the two terms in equation~(\ref{eq:P}), it easy to see that
\be
\label{eq:ratio}
   \frac{aT^4}{(\rho/m_p)kT}=\frac{\pi^2}{15}\frac{\theta^3}{\thd^3}.
\ee
Hence, the baryonic pressure gets dominant at $\theta\approx\thd$ [and at 
even lower temperatures, $\theta<(\pi/4)Y_e^{4/3}\thd$, the pressure of 
degenerate electrons 
$P_{\rm deg}=(3/4)(\pi/3)^{4/3}\hbar c(Y_e\rho/m_p)^{4/3}$ takes
over]. Our goal is to derive an upper bound on $\theta/\thd$ and
therefore we consider $\theta\simgt\thd$ with $P\approx aT^4$.
Constraint~(\ref{eq:constr}) then reads
\be
   \frac{\theta}{\thd}\simlt 2\rho_{11}^{-1/12}
     \left(\frac{r}{3r_g}\right)^{-1/4},
\ee
which also gives a lower bound on the electron chemical potential
(eq.~\ref{eq:mu_e}),
\be
   \frac{\mu}{Q}\simgt 5Y_e\rho_{11}^{1/2}
        \left(\frac{r}{3r_g}\right)^{1/2}.
\ee
We know from \S~2.1 that the condition $Y_e<0.5$ reads $\mu>Q/2$ for 
$\nu$-transparent matter and $\mu>Q$ for $\nu$-opaque matter.
We now see that any engine of size $r$ and density
$\rho>10^{10}(r/3r_g)^{-1}$~g/cm$^3$ satisfies this condition and
tends to an equilibrium $Y_e<0.5$. This is evidently the case in models
of neutron star mergers (Ruffert \& Janka 1999), as well as magnetized
neutron stars. The collapsar scenario (MacFadyen \& Woosley 1999) invokes 
a relatively low-density accretion flow, and here neutronization is 
questionable.  We therefore shall study accretion flows in more detail. 
We also need to check whether the equilibrium $Y_e$ is achieved on the 
accretion timescale.

\subsubsection{GRB accretion flows}

All accretion models of GRBs invoke rotation that creates a funnel
along which the fireball can escape. The accretion flow is viewed as a
rotating disk maintained in hydrostatic balance in the vertical direction,
which gradually spirals to the central black hole.
A standard model assumes turbulent viscosity in the disk with a stress tensor
$W_{r\phi}=\alpha c_s^2$, where $c_s=(P/\rho)^{1/2}$ is the isothermal sound
speed and $\alpha=0.01-0.1$ (Balbus \& Hawley 1998). Same stress tensor can
be described in terms of a viscosity coefficient $\nu=(2/3)\alpha c_s H$,
where $H$ is the half-thickness of the disk.

The standard disk theory gives the velocity of accretion at radius $r$
\be
 u^r=\frac{W_{r\phi}}{\OmK rS},
  \qquad S(r)=1-\left(\frac{\rin}{r}\right)^{1/2},
\ee
where $\OmK(r)=(GM/r^3)^{1/2}$ is the angular
velocity of Keplerian rotation, $M$ is the black hole mass, and
$\rin$ is the inner boundary of the disk (the marginally stable orbit);
$r_{\rm in}=3r_g$ for non-rotating and $r_{\rm in}=r_g/2$ for extremely 
rotating black holes.\footnote{We give here simple Newtonian estimates 
and trace the hole spin only through its effect on $r_{\rm in}$; other
relativistic corrections are modest and weakly affect the results.}
$S(r)$ varies from $0$ to $1$, and it equals $0.5$ at a characteristic 
radius $r=4r_{\rm in}$ where the integrated dissipation rate peaks
($4\pi Hr^2\dot{q}^+\propto S/r$, see eq.~[\ref{eq:qplus}] below).
One expects rotating black holes in GRBs, and $r\approx 3r_g$ is a 
reasonable characteristic radius.

The vertical hydrostatic balance reads $c_s=H\OmK$, which we use to write
the accretion time as
\be
\label{eq:ta}
   t_a(r)=\frac{r}{u^r}=
   \frac{S}{\alpha\OmK}\left(\frac{H}{r}\right)^{-2}=2.9\times 10^{-3}S
   \left(\frac{2H}{r}\right)^{-2}\left(\frac{\alpha}{0.1}\right)^{-1}
   \left(\frac{r}{3r_g}\right)^{3/2}\left(\frac{M}{M_\odot}\right){\rm ~s}.
\ee
We neglect here the disk gravity, which is a valid approximation as long as
$\dM t_a\ll M$ where $\dM$ is the accretion rate. The typical $t_a$ is much
shorter than the burst duration, and accretion is viewed as a quasi-steady
process. It can power a relativistic outflow (``fireball'') with luminosity
\be
  L=\epsilon_f\dM c^2\approx 10^{51}\left(\frac{\epsilon_f}{0.01}\right)
        \left(\frac{\dM}{10^{32}{\rm g~s}^{-1}}\right) {\rm ~erg~s}^{-1},
\ee
where $\epsilon_f$ is the efficiency of $\dM c^2$ conversion into a fireball,
which is below the net efficiency of accretion ($\epsilon\sim 0.1$).
Most of the accretion energy is either carried away by neutrinos
or advected by the accretion flow (in case of small neutrino losses).

The disk baryonic density is given by
\be
\label{eq:rho_}
  \rho =\frac{\dM t_a}{4\pi r^2 H}\approx 6.6\times 10^{10}S \dM_{32}
     \left(\frac{2H}{r}\right)^{-3}\left(\frac{r}{3r_g}\right)^{-3/2}
     \left(\frac{\alpha}{0.1}\right)^{-1}\left(\frac{M}{M_\odot}\right)^{-2}
    {\rm ~g~cm}^{-3},
\ee
and it is heated viscously with rate
\be
\label{eq:qplus}
  \dot{q}^+=\frac{3\dM\OmK^2}{8\pi H}S \approx 5.1\times 10^{33} S \dM_{32}
  \left(\frac{2H}{r}\right)^{-1}\left(\frac{r}{3r_g}\right)^{-4}
   \left(\frac{M}{M_\odot}\right)^{-3} {\rm ~erg~cm}^{-3}{\rm s}^{-1}.
\ee

The flow has a huge optical depth for radiation and the radiation is
trapped --- its diffusion is negligible on the accretion timescale. The only
cooling mechanism of the flow is neutrino
emission which becomes efficient at $\dM\simgt 10^{31}$~g~s$^{-1}$ and small
$r$ (e.g. Popham, Woosley, \& Fryer 1999; Narayan, Piran, \& Kumar 2001;
Kohri \& Mineshige 2002). An upper bound on the temperature is derived from
the assumption that the neutrino cooling is absent. Then the flow does not
loose the dissipated energy and instead traps it and advects. Its energy
density can be estimated as
$\wmax\approx\dot{q}^+t_a\approx(3/8)(r_g/r)\rho c^2$
(we use eqs.~[\ref{eq:ta}] and [\ref{eq:qplus}] with $S=0.5$).
The internal energy of such a hot advective flow is dominated by radiation
and $e^\pm$, so
\be
\label{eq:wmax}
   \wmax=\frac{11}{4}a\Tmax^4\approx \frac{3r_g}{8r}\rho c^2,
\ee
where factor $11/4$ accounts for the contribution of relativistic weakly
degenerate $e^\pm$ (neutrino contribution can further increase this factor
if the neutrinos are reabsorbed, and then $\Tmax$ will be slightly lower).
Equation~(\ref{eq:wmax}) yields
\be
\label{eq:Tmax}
   k\Tmax\approx 13\rho_{11}^{1/4}
          \left(\frac{r}{3r_g}\right)^{-1/4} {\rm ~MeV}.
\ee
The actual temperature can be significantly lower if the neutrino cooling
is significant.
The main cooling process is $e^\pm$ capture on nucleons:
$e^-+p\rightarrow n+\nu$ and $e^++n\rightarrow p+\bar{\nu}$, which also 
shapes $Y_e$ as we discussed in \S~2.1.
We now evaluate the characteristic accretion rate $\dMeq$ above which the 
$e^\pm$ capture is rapid enough to establish an equilibrium $Y_e$.

The equilibrium $Y_e$ is achieved
when the flow has emitted one neutrino per nucleon.
It is easy to see that disks with efficient neutrino cooling always reach 
the equilibrium. Indeed, the mean energy of the emitted neutrinos,
$\overline{E_\nu}\simlt 5kT$, is below the liberated accretion energy per
nucleon, $E_n=100-300$~MeV, and an efficient cooling implies that more than
one neutrino per nucleon is produced. Thus, $\dMeq$ should be looked for in 
the inefficient (advective) regime with $T\approx\Tmax$. Such a flow is only 
mildly degenerate, and the rates of $e^\pm$ capture read (in zero order in 
$\mu/\theta$)
\begin{eqnarray}
\label{eq:rates}
 \dn_{e^-p}\approx 1.5\times 10^{-2}n_p\theta^5 {\rm ~cm}^{-3}{\rm s}^{-1},\\
\label{eq:rates+}
 \dn_{e^+n}\approx 1.5\times 10^{-2}n_n\theta^5 {\rm ~cm}^{-3}{\rm s}^{-1}.
\end{eqnarray}
The neutronization timescale is $t_n=n_p/\dn_{e^-p}\approx 70\theta^{-5}$~s,
and it should be compared with the accretion timescale $t_a$ 
(eq.~\ref{eq:ta}),
\be
  \frac{t_n}{t_a}\approx 70
   \frac{\alpha\OmK}{S\theta^5}\left(\frac{H}{r}\right)^2.
\ee
We substitute $T=\Tmax$, take into account the hydrostatic balance
\be
 \frac{H}{r}=\left(\frac{\wmax}{3\rho}\right)^{1/2}\frac{1}{\OmK r}
            =\frac{1}{2},
\ee
and use equation~(\ref{eq:rho_}) with $S=0.5$ to get
\begin{eqnarray}
  \frac{t_n}{t_a}\approx
   1.7\times 10^{-2} \dM_{32}^{-5/4}\left(\frac{r}{3r_g}\right)^{13/8}
  \left(\frac{\alpha}{0.1}\right)^{9/4}\left(\frac{M}{M_\odot}\right)^{3/2}.
\end{eqnarray}
We conclude that disks with
\be
\label{eq:dMeq}
  \dM>\dMeq=3.8\times 10^{30}\left(\frac{r}{3r_g}\right)^{13/10}
  \left(\frac{\alpha}{0.1}\right)^{9/5}\left(\frac{M}{M_\odot}\right)^{6/5}
  {\rm ~g~s}^{-1}
\ee
have $t_n<t_a$ and hence approach the equilibrium $Y_e$. Radius $r$ entered 
this expression in the 13/10 power, which implies that the characteristic 
$\dMeq$ depends significantly on the black hole spin: the characteristic 
radius (where viscous dissipation peaks) decreases from $r\sim 10r_g$ to 
$r\sim r_g$ as the hole spin increases from zero to a maximum value.

The equilibrium $Y_e$ is below $0.5$ if the flow temperature is below $T_n$ 
that was calculated in \S~2.1 (eqs.~[\ref{eq:Tn_tr}] and [\ref{eq:Tn_op}]).
It is instructive to compare $T_n$ with the maximum accretion temperature 
$\Tmax$ (eq.~\ref{eq:Tmax}),
\be
\label{eq:TnTmax}
 \frac{\Tn}{\Tmax}=\kappa\rho_{11}^{1/4}
        \left(\frac{r}{3r_g}\right)^{1/4},
\ee
where $\kappa=2.5$ for $\nu$-transparent and $\kappa=1.8$ for $\nu$-opaque 
flows. The condition for a neutron-excess, $T<T_n$, is most difficult to 
satisfy in low-$\dM$ flows where the neutrino cooling is inefficient and 
$T\approx\Tmax$. Such flows are $\nu$-transparent, so that $\kappa=2.5$ 
should be used in equation~(\ref{eq:TnTmax}). Substituting 
equation~(\ref{eq:rho_}), $T=\Tmax$, and $H/r=1/2$, we find that 
$\Tmax<\Tn$ if
\be
\label{eq:dMn}
 \dM>\dM_n=7.6\times 10^{30}
    \left(\frac{r}{3r_g}\right)^{1/2}\left(\frac{\alpha}{0.1}\right)
    \left(\frac{M}{M_\odot}\right)^2{\rm ~g~s}^{-1}.
\ee
The ``neutronization'' accretion rate $\dM_n$ is comparable to $\dMeq$.
Plausible $\dM$ in GRB accretion flows are $10^{32}$~g/s and higher, and
they should have a neutron excess.

\subsection{De-neutronization in the fireball?}

Are outflows from neutron-rich engines also neutron-rich? The fireball picks
up baryons from the surface of the central engine, and the surface density 
is relatively low. $Y_e$ might change while the matter is escaping into a
fireball.

Let us consider fireballs produced by accretion disks.
The disk is turbulent (the turbulence is the source of viscosity that is
responsible for accretion), and it is mixed in the vertical direction on the 
sound-crossing timescale $\tmix\approx H/c_s$ (the turbulent velocity is 
somewhat smaller than $c_s$, however, the thickness of surface layers is 
also smaller than $H$, and $\tmix\approx H/c_s$ is about right). 
Given the hydrostatic balance, $H=c_s/\OmK$, one gets
\be
 \tmix\approx \OmK^{-1}.
\ee
The turbulent material circulates rapidly up to
the surface and back to the interior of the disk, and a small portion of
it can (also rapidly) escape in each circulation.
The escape timescale of the fireball is comparable to $\OmK^{-1}$.
As an element of matter elevates to the surface its $Y_e$ would increase if
it adjusted instantaneously to a new equilibrium value. However, the time
$\OmK^{-1}$ the element has before it sinks back into the disk (or escapes)
can be too short for the adjustment. Then $Y_e$ of the escaping matter
corresponds to $\rho$ and $T$ {\it inside} the disk, where it has spent
almost all the time before the sudden escape.

To check this picture, let us evaluate the timescale of ``de-neutronization''
of an initially neutron rich material that has suddenly expanded into a
low-density, hot fireball.
Neutrons tend to convert back into protons via two charged current reactions:
$e^+$ capture and $\nu$ absorption ($\beta$-decay is slow and negligible).
The fireball has temperature $\theta>1$ and non-degenerate electrons,
$\mu\ll \theta$. Equation~(\ref{eq:rates+}) then gives the $e^+$ capture
timescale
\be
   \tc=\frac{n_n}{\dn_{e^+n}}\approx\frac{70}{\theta^5} {\rm ~s},
\ee
\be
   \tc\OmK=\left(\frac{kT}{8{\rm ~MeV}}\right)^{-5}
   \left(\frac{r}{3r_g}\right)^{-3/2}\left(\frac{M}{M_\odot}\right)^{-1}.
\ee
For fireball temperatures up to 8~MeV the $e^+$-capture is slow compared
to $\OmK^{-1}$ and does not affect $Y_e$ of the escaping material.

The $\nu$ absorption by neutrons is potentially more important for
de-neutronization. The cross section of this reaction is
\be
   \sigma_a(\omega)=2.4\times 10^{-44}\left[1-f_-(\omega+Q)\right]
      (\omega+Q)^2\left[1-\frac{1}{(\omega+Q)^2}\right]^{1/2} {\rm ~cm}^2
             \approx 2.4\times 10^{-44}\omega^2 {\rm ~cm}^2,
\ee
and the corresponding rate of de-neutronization is
\be
\label{eq:zzz}
   \dn_n=-cn_n\int_0^\infty n_\omega\sigma_a\dd \omega,
\ee
where $n_\omega=\dd n_\nu/\dd\omega$ is the spectrum of the neutrino number
density. This gives the timescale of de-neutronization,
\be
  t_\nu=\frac{n_n}{|\dot{n}_n|}
   =\frac{1.4\times 10^{33}}{n_\nu\overline{\omega^2}} {\rm ~s},
\ee
where bar signifies an average over a neutrino spectrum.
An upper bound on the neutrino density outside the disk is given by
\be
   n_\nu<\frac{3\dM \OmK^2 S}{8\overline{\omega} m_ec^3},
\ee
which states that the energy flux of the (electron) neutrino from the two
faces of the disk, $2F_\nu\approx 2n_\nu\overline{\omega}m_ec^3/\pi$,
is smaller than the accretion energy released per unit time per unit area of
the disk, $F=(3/4\pi)GM\dM S/r^3$. Thus, we get (with $S\sim 0.5$)
\be
\label{eq:tnu}
  t_\nu\OmK > 10^2\frac{\overline{\omega}}{\overline{\omega^2}}
  \dM_{32}^{-1}\left(\frac{M}{M_\odot}\right)
  \left(\frac{r}{3r_g}\right)^{3/2}.
\ee
The mean $\omega$ and $\omega^2$ are determined by the state of material
that emits the neutrinos. If the disk is $\nu$-transparent, the neutrinos
have the spectrum (see eq.~[\ref{eq:e-p}])
\be
\label{eq:nuom}
 n_\omega\propto (\om+Q)^2\left[\exp\left(\frac{\omega+Q-\mu}{\theta}\right)
   +1\right]^{-1}\left[1-\frac{1}{(\om+Q)^2}\right]^{1/2}\om^2
  \approx \om^4\left[\exp\left(\frac{\omega}{\theta}\right)+1\right]^{-1},
\ee
where $\theta$ is temperature inside the disk, and $Q$ and $\mu$
have been neglected compared to the typical $\omega=\overline{\omega}$.
This gives
$\overline{\omega}=(31/6)[\zeta(6)/\zeta(5)]\theta\approx 5.07\theta$ and
$\overline{\omega^2}=(63/2)[\zeta(7)/\zeta(5)]\theta^2\approx 30.6\theta^2$.

The optical depth of the disk for neutrino absorption by neutrons
is\footnote{Absorption makes a major contribution to the disk
opacity for neutrinos.}
\be
\label{eq:tau_nu}
   \tau_a\approx \dM_{32}^{3/2}\left(\frac{\omega}{5\theta}\right)^2
    \left(\frac{T}{\Tmax}\right)^{-5}
    \left(\frac{r}{3r_g}\right)^{-7/4}\left(\frac{\alpha}{0.1}\right)^{-3/2}
    \left(\frac{M}{M_\odot}\right)^{-2}.
\ee
Here we assumed $T>\Td$, so that $Y_e$ is not much below 0.5 and 
$2H/r=(T/\Tmax)^2$ ($P\propto T^4$, see eqs.~[\ref{eq:P}] and 
[\ref{eq:ratio}]). From equations~(\ref{eq:tnu}) and (\ref{eq:tau_nu}) one 
can see that $\nu$-transparent disks ($\tau_\nu<1$) have $t_\nu\OmK>1$ for 
$\alpha>0.01$, and hence their fireballs do not experience any significant 
absorption of neutrinos.

High-$\dM$ disks with neutrino luminosity $L_\nu >10^{53}$~erg/s create a
sufficiently dense bath of neutrinos that can impact $Y_e$ of the ejected 
fireball. High-$\dM$ disks are $\nu$-opaque and the emitted neutrinos 
can be approximated by a blackbody spectrum with 
$\overline{\omega}\approx (7/2)[\zeta(4)/\zeta(3)]\theta_\nu\approx
3.15\theta_\nu$ and 
$\overline{\omega^2}\approx 15[\zeta(5)/\zeta(3)]\theta_\nu^2
\approx 12.9\theta_\nu^2$, where $\theta_\nu$ is the
temperature of the $\nu$-photosphere in the disk. The disk also
emits $\bar{\nu}$ from a corresponding $\bar{\nu}$-photosphere.
Both $\nu$ absorption by neutrons and $\bar{\nu}$ absorption by protons
occur at the base of the fireball and
set a new equilibrium $Y_e$ such that the rates of $\nu$ and
$\bar{\nu}$ absorptions are equal. One can expect the new $Y_e$ to be below
$0.5$ (see Qian et al. 1993, where a similar problem is discussed in the 
context of supernova engines, and Pruet, Fuller, \& Cardall 2001).
This expectation is based on two facts:
(1) the number densities of $\nu$ and $\bar{\nu}$ are approximately equal,
$|n_\nu-n_{\bar{\nu}}|\ll n_\nu$, --- otherwise they carry away too large
a leptonic number from the disk, and (2) the absorption cross section is
proportional to neutrino energy squared (in the limit $\omega\gg Q$,
which is valid for the hot high-$\dM$ disks).
It gives the equilibrium $n_n/n_p\approx \theta_{\bar{\nu}}^2/\theta_\nu^2$
where $\theta_\nu$ and $\theta_{\bar{\nu}}$ are temperatures of the $\nu$ 
and $\bar{\nu}$ photospheres. One expects $\theta_\nu<\theta_{\bar{\nu}}$
because the $\nu$-photosphere is likely closer to the disk surface
(the neutronized disk is more opaque for $\nu$) and then $n_n/n_p>1$.
This argument assumes, however, that the disk temperature decreases toward 
its surface. It may not hold if the dissipation rate peaks at the surface 
and makes it hotter than the interior of the disk.



\section{Nucleosynthesis}

The ejected fireball has a low baryon loading and a high temperature,
and its nucleons are initially in the free state. As expansion proceeds,
the fireball cools adiabatically and, when its temperature decreases to
$kT_*\sim 100$~keV, fusion reactions shape the nuclear composition like they
do during the
primordial nucleosynthesis in the Universe (Wagoner, Fowler, \& Hoyle 1967).
In both cases, we deal with an expanding blackbody fireball with initially
free nucleons, and the following three parameters control the outcome
of nucleosynthesis: (1) photon-to-baryon ratio $\phi=n_\gamma/n_b$,
(2) expansion timescale $\tnuc$ at the time of nucleosynthesis, and
(3) n/p-ratio prior to the onset of nucleosynthesis.
In the Universe, $\phi\approx 3\times 10^{9}$, $\tnuc\approx 10^2$~s, and
$n_n/n_p\approx 1/7$. Below we formulate the nucleosynthesis problem in GRBs
and give a qualitative comparison with the big bang. Then we
make detailed calculations of nuclear reactions in GRB fireballs.

\subsection{Fireball model}

A customary GRB model envisions a central engine that ejects baryonic matter
at a rate $\dM_b$~[g/s], thermal energy at a rate $\Lth\gg\dM_b c^2$, and
magnetic energy at a rate $\LP$. The produced fireball expands and
accelerates as its internal energy is converted into bulk kinetic energy.
The fireball is likely to carry strong magnetic fields and the Poynting
luminosity $\LP$ may be higher than the thermalized luminosity. In this case,
the magnetic fields prolong the acceleration after all the thermal energy has
been converted into bulk expansion. A maximum Lorentz factor of the fireball
can be estimated as $\Gsat=(\Lth+\LP)/\dM_b c^2$, and it is at least $10^2$
for GRB explosions.

The nucleosynthesis occurs when the fireball temperature drops to about
$kT_*\sim 100$~keV at a radius $\Rnuc\sim 10^7-10^9$~cm.
The timescale of expansion to this radius, $R_*/c$, is shorter than the
duration of the engine activity. Therefore, in the nucleosynthesis
calculations, the fireball can be modeled as a quasi-steady outflow.

Let the outflow expand in an axisymmetric funnel with a cross section
\be
\label{eq:S}
   S(R)=S_0\left(\frac{R}{r_0}\right)^{\psi}.
\ee
For example, $\psi=2$ for a radial funnel (and also for a spherically 
symmetric explosion), and $\psi=1$ for a parabolic funnel which may develop 
in a collapsing progenitor of the GRB (M\'esz\'aros \& Rees 2001).
The outflow is a relativistic ideal fluid with baryon density $\rho$,
pressure $P$, and energy density $w=3P\gg\rho c^2$; all these
magnitudes are measured in the fluid rest frame.
In spherical coordinates $x^i=(t,R,\theta,\phi)$ the outflow has 4-velocity
$u^i=\dd x^i/\dd\tau=(u^t,u^R,u^\theta,u^\phi)$, where $\tau$ is proper time.
We assume $u^\phi=0$ and $Ru^\theta\ll u^t,u^R$. The latter assumption is
satisfied at all radii for a radial explosion ($u^\theta=0$) and at 
$R\gg r_0$ for a collimated explosion ($\psi<2$).

The outflow dynamics is governed by the
conservation laws $\nabla_i(\rho u^i)=0$ and $\nabla_i(T^i_k)=0$, where
$T^i_k=u^iu_k(w+P)+P\delta^i_k$ is the stress-energy tensor.
The electromagnetic tensor is not included here, which greatly simplifies 
the problem (we are interested in the early hot stage when the expansion is 
likely driven by the thermal pressure even in the presence of strong fields). 
Then the baryon and energy conservation laws read
\begin{eqnarray}
\label{eq:laws}
  S\rho u^R=\dM_b, \qquad S(w+P)u^tu^R=\Lth.
\end{eqnarray}
The high Lorentz factor of the expansion $\Gamma=u^t\gg 1$ implies
$u^R/c\approx\Gamma$. Equations~(\ref{eq:laws}) then yield
\begin{eqnarray}
\label{eq:w}
  \rho=\frac{\dM_b}{S\Gamma c}, \qquad
  \frac{w}{\rho}=\frac{3}{4}\frac{\Lth}{\dM_b\Gamma}.
\end{eqnarray}
We assume that the outflow does not exchange mass or energy with the
surroundings, i.e., $\dM_b(R)={\rm const}$ and $\Lth(R)={\rm const}$.
Equation~(\ref{eq:w}) then gives $w\propto\rho/\Gamma$ while
the first law of thermodynamics $\dd(w/\rho)=-P\dd(1/\rho)$ gives
$w\propto \rho^{4/3}$. Excluding $\rho$ from these two relations one gets
$w^{1/4}\propto 1/\Gamma$ and
\be
\label{eq:T}
    T\Gamma=T_0,
\ee
where $T_0$ is a constant. Equation~(\ref{eq:T}) is strictly valid at
temperatures $kT\ll m_ec^2=511$~keV where the $e^\pm$ energy density can be
neglected. At small $\Gamma$ (where $kT\simgt m_ec^2$), the inclusion of
$e^\pm$ reduces $T$ by a modest factor $(11/4)^{-1/4}$.

Equation~(\ref{eq:laws}) implies $w=aT^4\propto (\Gamma^2S)^{-1}$
and, combining with equation~(\ref{eq:T}), one gets
\be
\label{eq:Gam}
   \Gamma=\frac{T_0}{T}=\left(\frac{S}{S_0}\right)^{1/2}
         =\left(\frac{R}{r_0}\right)^{\psi/2}.
\ee
The outflow may be collimated already at its base into a solid angle
$\Omega_0=S_0/r_0^2<4\pi$ --- its transonic dynamics near the central engine
is complicated and unknown. With reasonable accuracy, the simple estimate
$\Lth\approx S_0aT_0^4c$ gives
\be
\label{eq:T0}
  kT_0\approx 1.2 \left(\frac{\Lth}{10^{51}{\rm ~erg~s}^{-1}}\right)^{1/4}
     \left(\frac{r_0}{3\times 10^6{\rm ~cm}}\right)^{-1/2}
     \left(\frac{\Omega_0}{4\pi}\right)^{-1/4}~{\rm MeV}.
\ee
The value of $T_0$ is most sensitive to the engine size $r_0$.
It can be as small as $3\times 10^5$~cm (if the outflow is powered by
a Kerr black hole via the Blandford-Znajek process) or as large as $10^7$~cm
(if the outflow is powered by an accretion disk).

A major parameter of the nucleosynthesis problem is the ratio of
photon density,
\be
  n_\gamma=\frac{30\zeta(3)w}{\pi^4kT}\approx \frac{w}{2.70kT},
\ee
to baryon density $n_b$. This ratio is constant during the
expansion,\footnote{It is proportional to entropy per baryon:
$s/k=(2\pi^4/45\zeta[3])n_\gamma/n_b\approx 3.602n_\gamma/n_b$ where $k$ 
is Boltzmann constant, and it must be constant in an adiabatic expansion.}
and its value is (cf. eqs.~[\ref{eq:w}] and [\ref{eq:T}])
\be
\label{eq:eta}
\phi=\frac{n_\gamma}{n_b}=\frac{3}{4}\frac{m_p}{\dM_b}\frac{\Lth}{2.7kT_0}
    =7.7\times 10^{4}
   \left(\frac{\etath}{300}\right)\left(\frac{kT_0}{1{\rm ~MeV}}\right)^{-1},
  \qquad \etath=\frac{\Lth}{\dM_b c^2}.
\ee
The typical $\phi$ is 5 orders of magnitude smaller compared to that of
the Universe.

The nucleosynthesis must occur during the acceleration stage of the fireball
because the acceleration ends at a low temperature $T=T_0/\Gsat\ll\Tnuc$.
The accelerated expansion is described in the comoving time by
$\dd\tau=\dd t/\Gamma\approx \dd R/c\Gamma$, which gives (we use
eq.~[\ref{eq:Gam}])
\be
\label{eq:tau}
  \tau(T)=\left\{\begin{array}{ll}
       \frac{r_0}{c}\log\left(\frac{T_0}{T}\right)        &  \psi=2,\\
 \frac{r_0}{c}\frac{2}{(2-\psi)}\left[
     \left(\frac{T}{T_0}\right)^{(\psi-2)/\psi}-1\right]  &  \psi< 2.
              \end{array}
     \right.
\ee
The time of nucleosynthesis
depends on the shape of the funnel: $\tau(\Rnuc)\approx 3 r_0/c$ for $\psi=2$
and $\tau(\Rnuc)\approx 2(T_0/\Tnuc)(r_0/c)$ for $\psi=1$. The timescale
of the density fall off is 3 times shorter than that of the temperature 
because $\rho\propto T^3$. Therefore a reasonable choice of the 
characteristic expansion timescale during nucleosynthesis is 
$\tnuc\approx r_0/c$ for $\psi=2$ and $\tnuc\approx(2r_0/3c)(T_0/\Tnuc)$ 
for $\psi=1$.

Note that $\Lth$, $\dM_b$, and $T_0$ enter the nucleosynthesis problem only 
in combinations that determine $\phi$ and $\tau(T)$, and play no other role.
For example, the initial temperature is not important as long as 
$kT_0>200$~keV.

\subsection{Simple estimates and comparison to the big bang}

First we evaluate the temperature $\Tnuc$ at which we expect the 
nucleosynthesis to begin. Let us remind why $k\Tnuc\approx 80$~keV in the 
big bang.  If the nuclear statistical equilibrium (NSE) were maintained, 
recombination of nucleons into $\alpha$-particles would occur at 
$kT\sim 200$~keV (e.g. Meyer 1994). The reason of the nucleosynthesis delay 
until $kT\approx 80$~keV is what is sometimes called ``deuterium bottleneck''.
Before fusing into $\alpha$-particles, the nucleons have to
form lighter nuclei --- deuterium, tritium, or $^3$He. At $kT>100$~keV,
these elements have very low equilibrium abundances, which implies a very 
long timescale for their fusion, and helium is not formed even
though it is favored by the NSE. In particular, the deuterium
abundance is suppressed by the very fast photodisintegration
$\gamma+d\rightarrow n+p$ that balances the opposite reaction
$n+p\rightarrow d+\gamma$ at the equilibrium value
\be
\label{eq:Tnuc}
   Y_d\approx 7\times 10^{-6}\phi^{-1} Y_nY_pT_9^{3/2}
   \exp\left(\frac{25.8}{T_9}\right),
\ee
(e.g. Esmailzadeh, Starkman, \& Dimopoulos 1991).
Here $Y_i=n_i/n_b$ is abundance of the $i$-th element and
$T_9=T/10^9{\rm K}=kT/86.17{\rm~keV}$. Nucleosynthesis starts when the
exponential wins the pre-exponential factor to give a noticeable
$Y_d\sim 10^{-3}$.  Substituting $\phi=3\times 10^9$, one finds that it 
happens at $kT_*\approx 80$~keV.

In GRB outflows, $\phi$ is 5 orders of magnitude smaller, and the NSE
favors the recombination of nucleons into $\alpha$-particles at a temperature
as high as $500$~keV.
For the same reason as in the big bang, the nucleosynthesis is delayed to
a lower $T_*$. Using equation~(\ref{eq:Tnuc}) with $\phi=10^5$, we find 
$k\Tnuc\approx 140$~keV, which is almost twice as high as the big bang $T_*$
--- it depends logarithmically on $\phi$.

At a given $T$, nuclear reaction rates scale as
$\dot{Y}\propto\rho\propto\phi^{-1}$ (here $Y$ is an element abundance).
The ratio of the expansion timescale $\tau_*$ to a reaction timescale, 
$\treac=Y/\dot{Y}$, behaves as $\tau_*/\phi$. This combination is $\sim 3-30$ 
times smaller in GRBs compared to the big bang, which may seem not a crucial
difference. In fact, the big bang nucleosynthesis was dangerously close to
a freezeout of the neutron capture reaction $n+p\rightarrow d+\gamma$,
without which nucleosynthesis cannot start. Therefore the difference by a
factor of 10 can be crucial and it is instructive to compare accurately the
capture timescale with the expansion timescale in GRBs.

The neutron capture rate varies slowly with temperature
(see Fig.~11 in Smith, Kawano, \& Malaney 1993), and near 100~keV it is
approximately $\dY_c\approx 2.5\times 10^4\rho Y_pY_n$~s$^{-1}$. The reaction
timescale is
\be
  \tau_c=\frac{\min\{Y_p,Y_n\}}{\dot{Y}_c}
        =\frac{1.2\times 10^{-9}\phi}{T_9^3\max\{Y_n,Y_p\}} {\rm ~s}.
\ee
It should be compared with the timescale of the density fall off, $\tau_*$.
For the big bang, $\tnuc/\tau_c\approx 30$. For GRBs, we get
\be
  \tau_c\approx 10^{-4}\left(\frac{\phi}{10^{5}}\right)
         \left(\frac{Y_n}{0.5}\right)^{-1}{\rm ~s},
\ee
\be
  \frac{\tnuc}{\tau_c}\approx \left\{\begin{array}{ll}
   \left(\frac{\phi}{10^{5}}\right)^{-1}
   \left(\frac{r_0}{3\times 10^6}\right)
   \left(\frac{Y_n}{0.5}\right) & \psi=2, \\
   \left(\frac{\phi}{10^{5}}\right)^{-1}
   \left(\frac{r_0}{3\times 10^6}\right)
   \left(\frac{Y_n}{0.5}\right)
   \left(\frac{T_0}{T_*}\right) & \psi=1.
    \end{array}
     \right.
\ee
One can see that, in a radial explosion ($\psi=2$), the neutron capture rate
is marginal for a successful nucleosynthesis. In a collimated explosion,
the ratio $\tnuc/\tau_c$ is higher by a factor of $T_0/T_*\sim 10$,
and nucleosynthesis is efficient.

An important difference between the GRBs and the big bang
is the n/p-ratio. In the big bang, $n_n/n_p=1/7$ ($Y_e=7/8$), which
leads to 25\% mass fraction of helium after the n-p recombination, while
75\% of mass remains in protons (and a tiny amount of other nuclei).
In GRBs, $n_n/n_p>$1 ($Y_e<0.5$), and there are leftover neutrons even if
all protons are consumed by helium production.
The minimum mass fraction of leftover neutrons is
\be
\label{eq:Xn}
  X_n=1-2Y_e.
\ee

\subsection{Detailed calculation}

\subsubsection{The code}

We will keep track of elements with mass numbers less than 5 (like the
big bang, the abundances of heavier nuclei are very small).
The six elements under consideration are neutrons, protons, $^2$H, $^3$H,
$^3$He, and $^4$He; they are denoted by $n,p,d,t,3,\alpha$,
respectively, and the photons are denoted by $\gamma$. Their abundances
are measured by $Y_i=n_i/n_b$ or mass fraction $X_i=A_iY_i$, where $n_i$ and
$A_i$ are the number density and the mass number of the i-th species.
The photon and matter densities are known functions of temperature,
\be
   n_\gamma=\frac{2}{\pi^2}\zeta(3)\left(\frac{kT}{\hbar c}\right)^3
    =2.02\times 10^{28}T_9^3{\rm ~cm}^{-3},
   \qquad \rho=3.39\times 10^4\phi^{-1} T_9^3{\rm ~~g~cm}^{-3}.
\ee

The evolution of nuclear composition is described by the set of equations
\begin{eqnarray}
\label{eq:Yi}
 \dot{Y_i} & = & \sum Y_kY_l[klij]-\sum Y_iY_j[ijkl],\\
\label{eq:Yn}
 \dot{Y}_n & = & -\dot{Y_d}-2\dot{Y_t}-\dot{Y_3}
                 -2\dot{\Ya}-\frac{Y_n}{\taub},\\
\label{eq:Yp}
 \dot{Y}_p & = & -\dot{Y_d}-\dot{Y_t}-2\dot{Y_3}-2\dot{\Ya}+\frac{Y_n}{\taub}.
\end{eqnarray}
Here dot signifies a derivative with respect to proper time $\tau$,
$[ijkl]=n_b\overline{\sigma v}_{ij\rightarrow kl}$ is the rate of reaction
$i+j\rightarrow k+l$, and all quantities are measured in the rest frame of
the outflow. It is sufficient to calculate the reaction rates
for deuterium, tritium, and helium isotopes $^3$He and $^4$He
($i=d,t,3,\alpha$ in eq.~\ref{eq:Yi}). Then $\dot{Y}_n$ and $\dot{Y}_p$
are found from the neutron and proton conservation laws (eqs.~[\ref{eq:Yn}]
and [\ref{eq:Yp}]). The latter include the $Y_n/\taub$ term --- the conversion
of neutrons into protons via $\beta$-decay with the mean lifetime of neutrons
$\taub=900$~s. The $\beta$-decay is negligible in GRBs and
we keep it for the code tests on big bang nucleosynthesis.

The sums in equation~(\ref{eq:Yi}) are
taken over all possible reactions with participation of the i-th nuclei.
Not all reactions are important (Smith et al. 1993). For example, reactions
that destroy $\alpha$-particles can be neglected as $Y_\alpha$ is far
below its equilibrium value.
We include the following reactions in the calculations:
$n+p\leftrightarrow d+\gamma$,
$n+3\leftrightarrow p+t$, $n+3\rightarrow \alpha+\gamma$,
$p+d\rightarrow 3+\gamma$, $p+t\rightarrow \alpha+\gamma$,
$d+d\rightarrow p+t$, $d+d\rightarrow n+3$,
$d+t\rightarrow n+\alpha$, $d+3\rightarrow p+\alpha$,
and take their rates from Smith et al. (1993)
and Esmailzadeh et al. (1991). We then have
\begin{eqnarray}
\nonumber
 \dot{Y_d} & = & Y_nY_p[npd\gamma]-Y_dY_\gamma[d\gamma np]-Y_pY_d[pd3\gamma]
           -2Y_d^2[ddn3]-Y_dY_3[d3p\alpha] \\
\label{eq:Yd}
       &   & -Y_dY_t[dtn\alpha]-2Y_d^2[ddpt], \\
\label{eq:Yt}
 \dot{Y_t}& = &Y_d^2[ddpt]+Y_nY_3[n3pt]-Y_dY_t[dtn\alpha]-Y_pY_t[ptn3]
            -Y_pY_t[pt\alpha\gamma],\\
\label{eq:Y3}
 \dot{Y_3}& = &Y_pY_d[pd3\gamma]+Yd^2[ddn3]+Y_pY_t[ptn3]-Y_dY_3[d3p\alpha]
            -Y_nY_3([n3pt]+[n3\alpha\gamma]),\\
\label{eq:Ya}
 \dot{\Ya}& = & Y_dY_t[dtn\alpha]+Y_dY_3[d3p\alpha]+Y_nY_3[n3\alpha\gamma]
            +Y_pY_t[pt\alpha\gamma].
\end{eqnarray}
The reaction rates are functions of $T$ and $\rho$, and the set of
equations is closed by equation~(\ref{eq:tau}) that relates $T$ and $\tau$.
We solve equations~(\ref{eq:Yn}-\ref{eq:Ya})
numerically with initial conditions $Y_p^0=Y_e$, $Y_n^0=1-Y_e$, $\Ya=0$, and
$\dot{Y}_d=\dot{Y}_t=\dot{Y}_3=0$ at an initial temperature $kT_0\sim 1$~MeV.
At high temperatures, the abundances of all elements except $n$ and $p$
are negligibly small. The abundances of deuterium,
tritium, and $^3$He are close to quasi-steady equilibrium at $kT>150$~keV,
i.e., the rates of their production $\dot{Y}^+$ and sink $\dot{Y}^-$
are much higher than the expansion rate $Y/\tau$, and hence
$\dY^+$ and $\dY^-$ almost balance each other. Where an element abundance
$Y$ approaches the equilibrium value, it is calculated by setting the net
$\dot{Y}=\dot{Y}^+ - \dot{Y}^-=0$. (Thus, one avoids numerical integration
where $\dot{Y}$ is a small difference of big numbers.) The code has been 
tested with the big bang nucleosynthesis. It successfully reproduced the
standard evolution of the cosmological nuclear composition.

\subsubsection{Results}

Figure~3 compares the nucleosynthesis in GRBs and the big bang (BB).
Parameters of the GRB fireball in this example are chosen as
$\phi=10^{5}$, $r_0=3\times 10^6$~cm, $\psi=2$, and $Y_e=0.5$.
As expected, GRB nucleosynthesis occurs at higher temperatures. The freezeout 
mass fractions of helium and deuterium are $X_\alpha\approx 0.16$ and 
$X_d\approx 0.03$, and about 81\% of mass remains in free nucleons.
Interestingly, the deuterium evolution is qualitatively different in GRBs:
$X_d$ increases monotonically and freezes out at the 3\% level which is 
almost 3 orders of magnitude higher than in the big bang (see also 
Pruet et al. 2002).  This difference is due to a high abundance of neutrons 
and a negligible $\beta$-decay. As a result, the deuterium production by 
the neutron capture reaction outweighes its burning rate and $X_d$ grows. 
Also, the freezeout happens quickly in the radial expansion 
($T\propto\exp[\tau/\tau_0]$ where $\tau_0=r_0/c$) and $X_d$ could not 
decrease much below 1\% even if the neutron capture reaction were completely 
switched off at, e.g., $kT<80$~keV.

\begin{figure}
\plotone{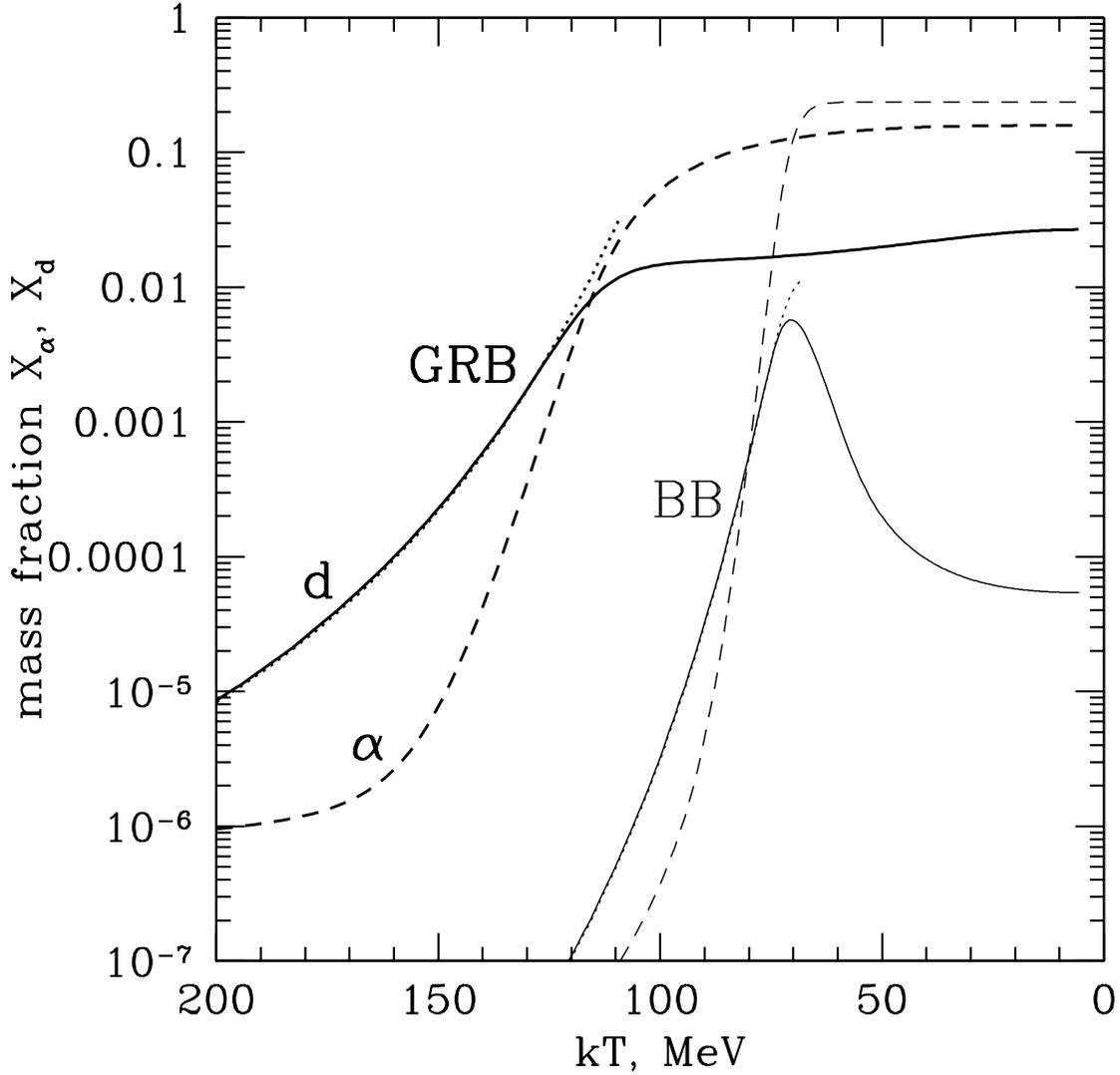}
\caption{Evolution of deuterium and helium abundances with temperature in an
expanding fireball. The GRB case is shown by thicker curves for a radial
explosion with $Y_e=0.5$, $\phi=n_\gamma/n_b=10^5$, and assuming a central
engine of size $r_0=3\times 10^6$~cm ($r_0$ sets the expansion timescale, 
cf.  eq.~[\ref{eq:tau}]). For comparison, the big bang (BB) nucleosynthesis 
is also shown (with $\phi=3\times 10^9$). The dotted curves display the 
equilibrium (production=sink) abundances of deuterium.
}
\end{figure}
\begin{figure}
\plotone{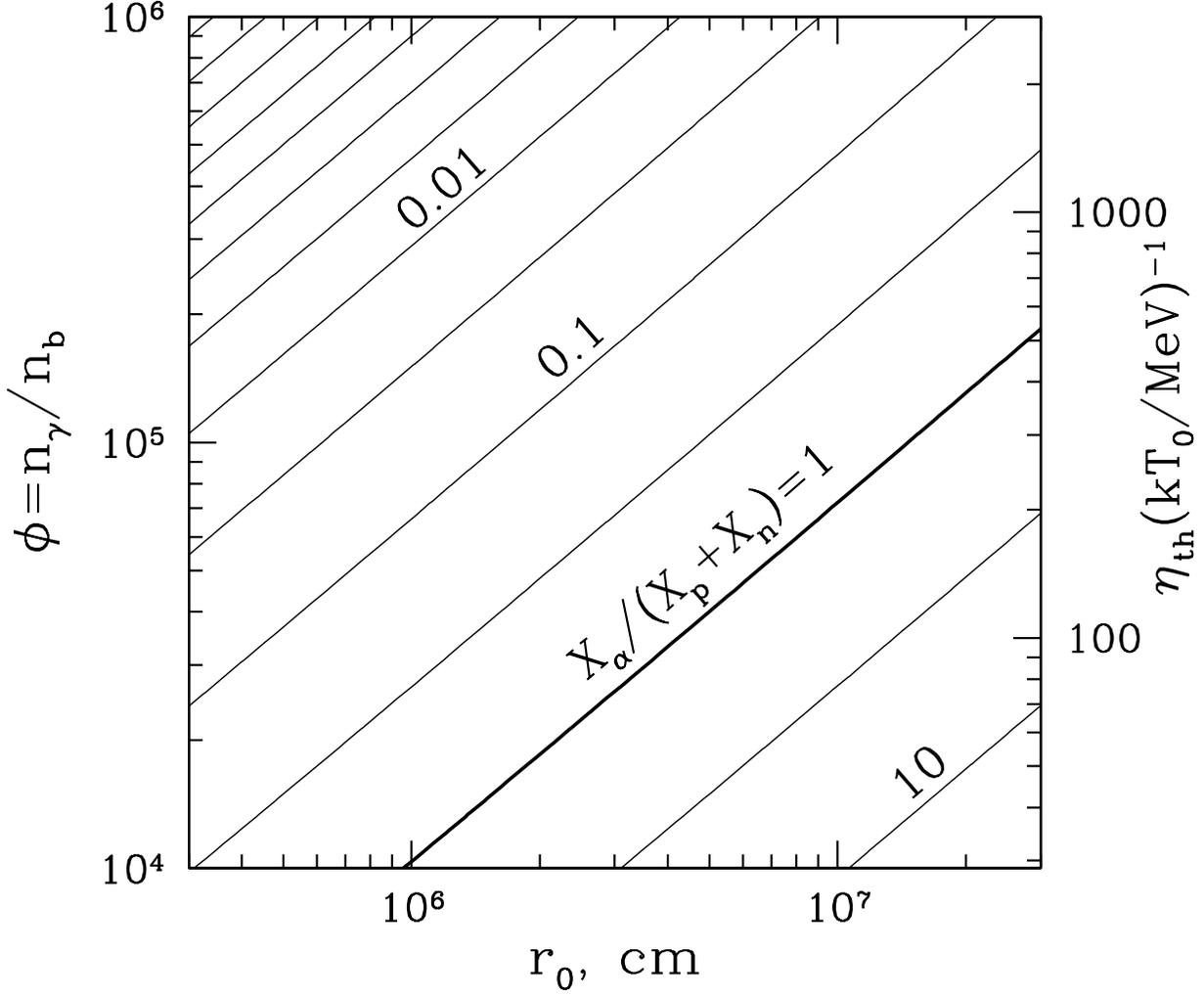}
\caption{Contours of freezeout ratio $f=X_\alpha/(X_p+X_n)$ on the
$r_0-\phi$ plane for radial explosions with $Y_e=0.5$. Right axis shows
$(\Lth/\dM_b c^2)(kT_0/{\rm MeV})^{-1}$, which is related to $\phi$ by
equation~(\ref{eq:eta}). If the fireball is not Poynting-flux dominated,
$\LP<\Lth$, its final Lorentz factor equals $\etath=\Lth/\dM_b c^2$.}
\end{figure}
\begin{figure}
\plotone{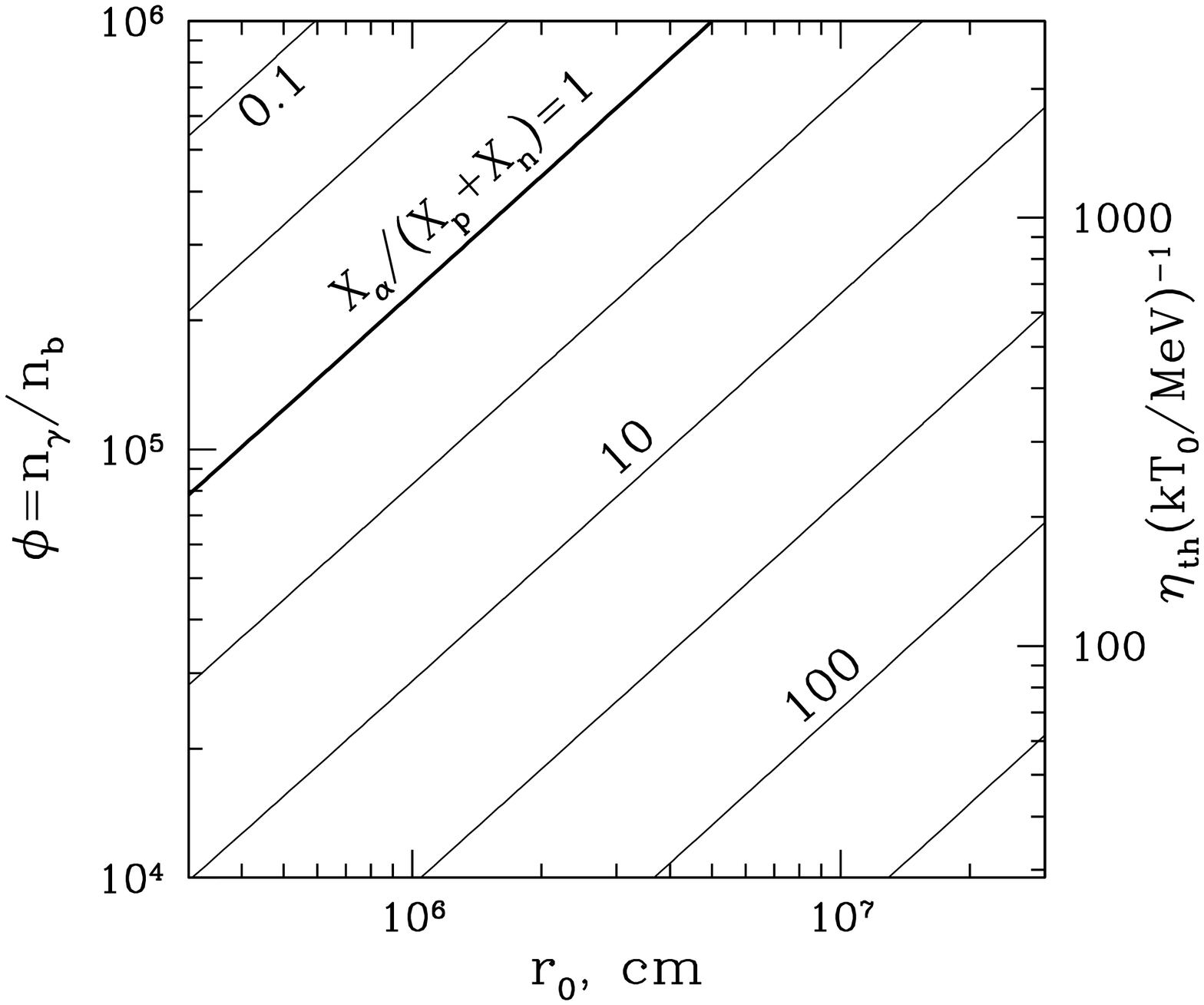}
\caption{Same as in Figure~4 but for parabolically collimated explosions.}
\end{figure}

The post-nucleosynthesis fireball is predominantly composed of free
nucleons and helium, and it would be useful to know how their freezeout ratio
\be
  f=\frac{X_\alpha}{X_n+X_p}
\ee
depends on the four parameters of the explosion $r_0$, $\phi$, $Y_e$, and 
$\psi$. Figure~4 shows $f(r_0,\phi)$ for radial explosions with $Y_e=0.5$. 
Contours $f={\rm const}$ on the $r_0-\phi$ plane are perfect straight 
lines $\phi\propto r_0^{-0.85}$ (or $r_0\propto \phi^{-1.18}$).
This implies that $f$ depends on combination $r_0\phi^{-1.18}$.
A similar situation takes place for collimated explosions (Fig.~5). 
The slope of all contours in Figure~5 is $0.90$ and $f$ can be viewed as a 
function of $r_0\phi^{-1.11}$. Calculations with $Y_e\neq 0.5$ give 
different values of $f$, however, contours $f(r_0,\phi)={\rm const}$ have 
exactly the same slope as in the case of $Y_e=0.5$.

The results of nucleosynthesis calculations can be
represented in a compact form if we define a new variable
\be
\label{eq:xi}
   \xi=\left\{\begin{array}{ll}
    0.2\left(\frac{\phi}{10^{5}}\right)^{-1.18}
   \left(\frac{r_0}{3\times 10^6}\right)  & \psi=2, \\
   7.6\left(\frac{\phi}{10^{5}}\right)^{-1.11}
    \left(\frac{r_0}{3\times 10^6}\right) & \psi=1.
    \end{array}
     \right.
\ee
Figure~6 shows the outcome of nucleosynthesis $f(\xi)$ for different $Y_e$.
Strikingly, $f(\xi)$ is practically identical for radial and collimated
explosions. Figure~6 gives a full description of helium
production in relativistic explosions: from this figure and
equation~(\ref{eq:xi}) one finds $f$ for given $\phi$, $r_0$, $Y_e$,
and $\psi=1,2$.

One can see that nucleosynthesis is most efficient in $Y_e=0.5$ models,
where $X_n^0=X_p^0$. In the most interesting region of parameters, where
$f$ changes from $f\ll 1$ to $f\gg 1$, one observes a simple behavior
\be
  f=\xi, \qquad 0.03<f<30,
\ee
for both radial and collimated explosions (this explains our choice of
constants 0.2 and 7.6 in eq.~[\ref{eq:xi}]). 

Fireballs with $Y_e\neq 0.5$
have an upper bound on the freezeout abundance of helium
$\max X_\alpha=2\min\{Y_e,1-Y_e\}$. For explosions with a neutron excess,
$Y_e<0.5$, this gives a  maximum $f$, 
\be
  \fmax=\frac{2Y_e}{1-2Y_e}.
\ee
With increasing $\xi$, the dependence $f\approx\xi$ switches to
$f=\fmax={\rm const}$ (Fig.~6).

\begin{figure}
\plotone{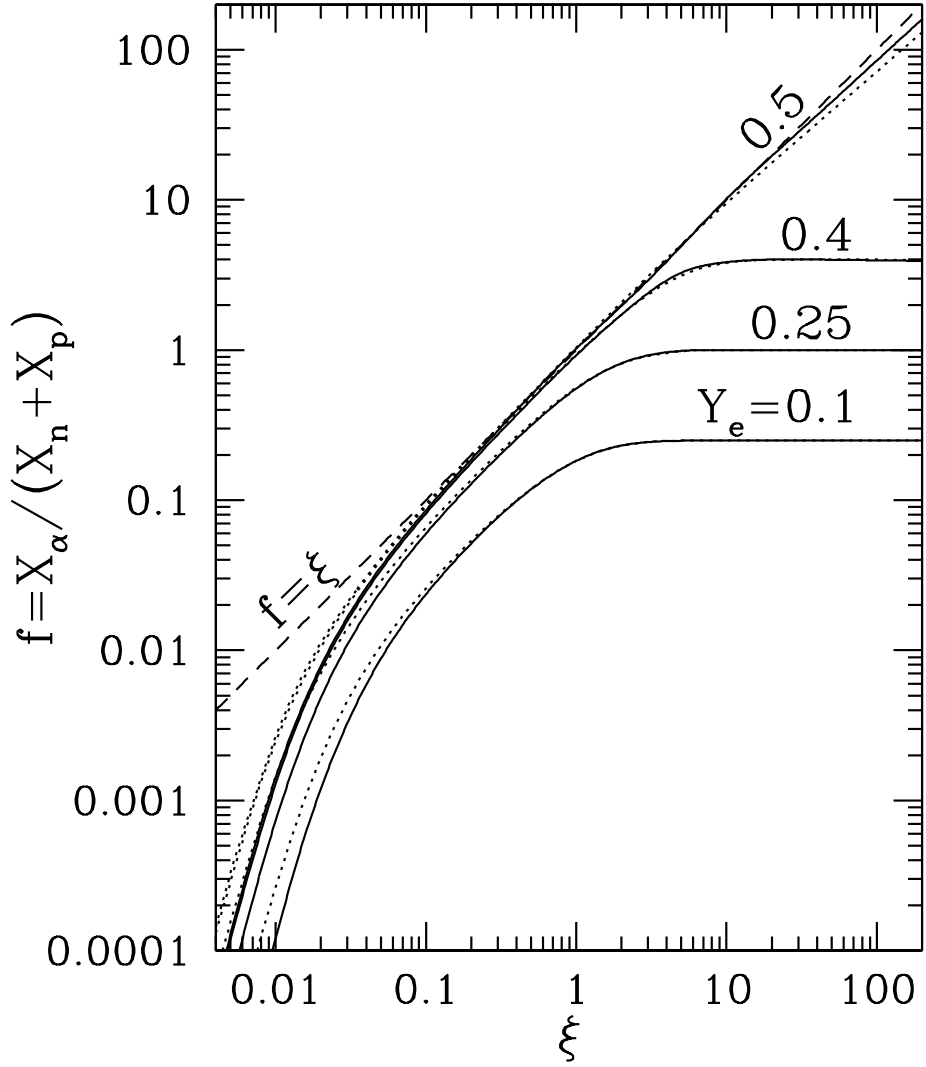}
\caption{Outcome of nucleosynthesis in relativistic explosions for $Y_e=0.1$,
0.25, 0.4, and 0.5. The ratio of mass fractions of helium and free nucleons,
$f=X_\alpha/(X_n+X_p)$, is shown as a function of $\xi$ defined in
equation~(\ref{eq:xi}). Dotted and solid curves display radial ($\psi=2$) and
collimated ($\psi=1$) explosions, respectively. Dashed line shows $f=\xi$, 
which well approximates the numerical results in the range $0.03<f<\fmax$ 
(if $Y_e$ is not much below 0.5).
}
\end{figure}

The post-nucleosynthesis fireball is dominated by $\alpha$-particles if 
$f>1$, which requires $Y_e>0.25$ and $\xi>1$. For example, in a radial 
explosion with $Y_e=0.5$ it requires 
$\phi<2.5\times 10^{4}(r_0/3\times 10^6)^{0.85}$, and a similar condition for
parabolic explosions reads $\phi<6.2\times 10^{4}(r_0/3\times 10^6)^{0.9}$.
The typical GRB parameters happen to be just marginal for a successful
nucleosynthesis: $f$ varies from $10^{-2}$ to $\fmax$ in the expected range 
of $\phi$ and $r_0$.



\section{Spallation}

Synthesized helium may be destroyed during the subsequent evolution of the
explosion. Spallation reactions can occur when an $\alpha$-particle collides
with another particle with a relative energy exceeding the nuclear binding 
energy. The fireball temperature at the acceleration stage is too low for 
such reactions, however, the collisions may become energetic if (1) there 
appears a substantial relative bulk velocity between the neutron and ion 
components or (2) internal shocks occur and heat the ion fireball to a high 
temperature, much above the blackbody value.

\subsection{Neutron-ion collisions during the acceleration stage}

Near the central engine, the neutron component of the fireball is well 
coupled to the ion component by elastic collisions with a small relative 
velocity $\tbet=\tilde{v}/c$ (Derishev et al. 1999, Bahcall \& M\'esz\'aros 
2000).  The collision cross section can be approximated as 
$\sigma_i=\sigma_0/\tbet$, where $\sigma_0\approx 3\times 10^{-26}$~cm$^2$ 
if the ions are protons (and $\sigma_0$ approaches $10^{-25}$~cm$^2$ for 
$\alpha$-particles).  In the fluid frame, the mean collisional time for 
neutrons is
\be
\label{eq:taucoll}
  \taucoll=\frac{1}{n_i\sigma_i \tbet c}=\frac{1}{n_i\sigma_0 c},
\ee
where $n_i$ is the ion density.
The ion density is a fraction of the total nucleon density $n_b$
which behaves as $n_b={\rm const}/\Gamma^3$ during the acceleration stage
(see eqs.~[\ref{eq:w}] and [\ref{eq:Gam}]).
The constant of proportionality can be expressed in terms of
$\etath=\Lth/\dM_b c^2$ and $\phi=n_\gamma/n_b$,
\be
 n_b=\frac{45^3\zeta^4(3)}{4\pi^{14}}\left(\frac{m_pc}{\hbar}\right)^3
     \frac{\etath^3}{\phi^4\Gamma^3}
    =5.61\times 10^{38}\frac{\etath^3}{\phi^4\Gamma^3} {\rm ~cm}^{-3}.
\ee
In the fixed lab frame, the ion fluid is accelerated by radiative or magnetic
pressure, and its Lorentz factor is doubled on timescale $t=R/c$. Neutrons
``miss'' the ion acceleration by
$\Delta\Gamma/\Gamma\approx\tcoll/t=\Gamma c\taucoll/R<1$ and have
a smaller Lorentz factor $\Gamma_n=\Gamma-\Delta\Gamma$.
The relative velocity of the neutron and ion components is
$\tbet=(\Gamma^2-\Gamma_n^2)/(\Gamma^2+\Gamma_n^2)
\approx (\Gamma-\Gamma_n)/\Gamma$, which is
\be
  \tbet=\frac{\tcoll}{t}=\frac{\Gamma}{Rn_i\sigma_0}
       \propto\frac{\Gamma^4}{R}.
\ee

The energy of neutron-ion collisions becomes sufficient for spallation
reactions if $\tbet$ exceeds $\tbsp\approx 0.1$. This can happen
at late stages of the fireball acceleration, at high $\Gamma$ but not
exceeding $\Gsat=L/\dM_b c^2=\etath+\etaP$ ($L=\Lth+\LP$ is the
total luminosity of the fireball that includes the Poynting flux).
In the case of $\Gamma(R)\approx R/r_0$ (radial explosion),
$\tbet$ reaches $\tbsp$ when $\Gamma$ reaches
\be
\label{eq:Gsp}
  \Gsp=0.3\Gsat\left(\frac{\Lth}{L}\right)
 \left(\frac{\phi}{10^{5}}\right)^{-4/3}\left(\frac{\tbsp}{0.1}\right)^{1/3}
 \left(\frac{r_0}{3\times 10^6}\right)^{1/3}
 \left(\frac{4n_i}{n_b}\right)^{1/3}.
\ee
Spallation takes place if $\Gsp<\Gsat$, which requires $\phi>\phisp$,
\be
  \phisp\approx 4\times 10^{4}\left(\frac{\Lth}{L}\right)^{3/4}.
\ee
We keep here $\Lth/L<1$ as the most uncertain parameter which may be much
below unity. Note also, that in the case of a non-radial explosion, $r_0$
should be replaced by $R/\Gamma$ in equation~(\ref{eq:Gsp}). Using
equation~(\ref{eq:eta}) it is easy to show that the condition $\phi>\phisp$ 
is equivalent to
\be
\label{eq:nal}
 \Gsat>160\left(\frac{kT_0}{\rm MeV}\right)\left(\frac{L}{\Lth}\right)^{1/4}.
\ee

Let $Y_n$ be abundance of neutrons that survived the nucleosynthesis, and
suppose that $\phi>\phisp$. The lifetime of $\alpha$-particles bombarded by
neutrons with $\tbet\approx\tbsp$ is
$\tlife\approx(\tcoll/Y_n)(\sigma_0/\ssp)\approx (\tbsp/Y_n)t$,
where $\ssp\approx\sigma_0$ is the spallation cross section (it is roughly
equal to the geometrical size of the nucleus
$\pi r_N^2\approx 5.3\times 10^{-26}A^{2/3}{\rm ~cm}^2$ with $A=4$ for 
$\alpha$-particles). Hence a modest $Y_n\sim 0.1$ should be sufficient for 
a significant spallation.

In case of a very low baryon loading (high $\Gsat$) the neutrons
decouple from the ions before the end of the acceleration stage and their
Lorentz factor $\Gamma_n$ saturates at $\Gdec<\Gsat$.
The relative velocity $\tbet$ approaches unity when the decoupling happens,
and the last neutron-ion collisions are energetic enough for pion production
which leads to multi-GeV neutrino emission (see Derishev et al. 1999,
Bahcall \& M\'esz\'aros 2000). The fireball Lorentz factor at this moment, 
$\Gdec$, is given by equation~(\ref{eq:Gsp}) with $\tbsp$ replaced by unity, 
i.e. $\Gdec=10^{1/3}\Gsp$. The decoupling takes place if $\Gdec<\Gsat$, 
which requires $\phi$ greater than
\be
  \phidec\approx 7\times 10^{4}\left(\frac{\Lth}{L}\right)^{3/4}.
\ee
The condition $\phi>\phidec$ is equivalent to
$\Gsat>300(kT_0/{\rm MeV})(L/\Lth)^{1/4}$.
The decoupling is always preceded by spallation.

The upper bound on the neutron Lorentz factor due to decoupling
is essentially determined by the rate of baryon outflow per unit solid
angle $\dM_\Omega$ [g/s], and it is useful to rewrite $\Gdec$ in terms of
$\dM_\Omega$. We substitute $n_i=n_b=(\dM_\Omega/R^2\Gamma m_p c)$ to
equation~(\ref{eq:taucoll}), and then the decoupling condition
$\Gamma\taucoll=R/c$ gives the maximum Lorentz factor of neutrons
\be
 \Gdec\approx\left(\frac{\sigma_0\dM_\Omega}{r_0m_pc}\right)^{1/3}
   \approx 300\left(\frac{\dM_\Omega}{10^{26}{\rm g~s}^{-1}}\right)^{1/3}
              \left(\frac{r_0}{3\times 10^{6}{\rm cm}}\right)^{-1/3}.
\ee

\subsection{Internal shocks}

The $\alpha$-particles can also be destroyed later on, when internal shocks 
develop in the fireball.  Lorentz factor $\Gsat$ fluctuates if the central 
engine is ``noisy'' during its operation $0<t<t_b\sim 1$~s. The fluctuations 
probably occur on timescales $r_0/c\approx 10^{-4}$~s and longer, up to $t_b$.
This leads to internal shocks (Rees \& M\'esz\'aros 1994). Internal 
dissipation of the velocity fluctuations may give rise to the observed GRB, 
and this picture is plausible because it easily accounts for the observed 
short-timescale variability in GRB light curves. The amplitude of the 
fluctuations is described by the dimensionless rms of the Lorentz factor, 
$A=\delta\Gsat/\Gsat$.  At modest $A\simlt 1$, the internal dissipation 
proceeds in a simple hierarchical manner (Beloborodov 2000). The internal 
collisions begin at
\be
\label{eq:R_0}
  R_0=\frac{2\Gsat^2}{A}\lambda_0=5.4\times 10^{11}A^{-1}
  \left(\frac{\Gsat}{300}\right)^2\left(\frac{\lambda_0}{3\times 10^6}\right)
  {\rm ~cm},
\ee
where $\lambda_0$ [cm] is a minimum length scale of the fluctuations, which 
is likely comparable to $r_0$. The shortest fluctuations are dissipated 
first, and at $R>R_0$ fluctuations with $\lambda>\lambda_0$ are dissipated 
in the hierarchical order.

\subsubsection{$n$-$\alpha$ collisions}

Suppose $\Gsat$ is small enough, so that the neutrons decouple from the ions
after the acceleration stage and have same $\Gamma_n=\Gsat$.
As soon as internal shocks develop in the ion component,
the neutrons begin to drift with respect to the ions. Their relative velocity
$\tbet\approx A$ is sufficient for spallation reactions if $A>0.1$
(and for $A\sim 1$ the pion production and multi-GeV neutrino
emission takes place, see M\'esz\'aros \& Rees 2000).

Let $L_\Omega=R^2n_bm_p c^3\Gsat^2$ be the fireball kinetic
luminosity per unit solid angle; then
\be
  n_b=\frac{L_\Omega}{m_pc^3R^2\Gsat^2}.
\ee
The $\alpha$-particles are destructed by neutrons with rate
$\dot{Y}_\alpha\approx Y_\alpha Y_n n_b\ssp c$ and their lifetime in the
fluid frame is $\tau_{\rm life}=(Y_nn_b\ssp c)^{-1}$,
\be
\label{eq:tlife}
  \taulife=\frac{m_pc^2R^2\Gsat^2}{\ssp Y_n L_\Omega}.
\ee
It should be compared with the timescale of side expansion in the fluid 
frame, $(R/c\Gsat)$. If $\taulife<(R/c\Gsat)$ at $R=R_0$ then most of the
$\alpha$-particles are destroyed by the internal shocks.
This condition reads
\be
\label{eq:drift}
  \Gsat<\left[\frac{AY_n\ssp L_\Omega}{2\lambda_0 m_pc^3}\right]^{1/5}
  \approx 300A^{1/5}Y_n^{1/5}\left(\frac{L_\Omega}{10^{52}}\right)^{1/5}
            \left(\frac{\lambda_0}{3\times 10^6}\right)^{-1/5}.
\ee

\subsubsection{$\alpha$-$\alpha$ collisions}

The $\alpha$-particles acquire random energy $(A^2/2)4m_pc^2$ in internal
shocks, which easily exceeds the spallation threshold. The condition
$\taulife<(R/c\Gsat)$ for efficient $\alpha$-$\alpha$ spallation is similar
to equation~(\ref{eq:drift}) (with $Y_n$ replaced by $Y_\alpha$). There is,
however, one more condition. The shocked $\alpha$-particles are cooled by 
Coulomb interactions with $e^-$ (or $e^+$) on a timescale $\tauCoul$, and 
an efficient spallation requires $\taulife<\tauCoul$ in addition to 
$\taulife<(R/c\Gsat)$.

The $e^\pm$ can be considered as targets at rest for the hot ions because
their radiative losses (synchrotron and/or inverse Compton) are rapid 
compared to the expansion rate. In the fluid frame, hot ions with mass 
$m_i$ and charge $Ze$ loose their random velocity $\tbet$ on timescale
(Ginzburg \& Syrovatskii 1964)
\be
\label{eq:tCoul}
  \tauCoul=\tbet\left(\frac{\dd\tbet}{\dd\tau}\right)^{-1}
          =\frac{2\tbet^3 m_i}{3Z^2\ln\Lambda\sT m_ec n_e},
\ee
where $n_e=n_-+n_+$ is the total density of $e^-$ and $e^+$, and
$\ln\Lambda\approx 20$ is Coulomb logarithm. The internal shocks give
$\tbet\approx A$ and, for $\alpha$-particles ($m_i=4m_p$, $Z=2$), we find
$\tau_{\rm life}=(n_\alpha\ssp c)^{-1}<\tauCoul$ if
\be
\label{eq:condit}
 \frac{n_e}{n_\alpha}<\frac{2A^3}{3\ln\Lambda}\frac{m_p}{m_e}\frac{\ssp}{\sT}
 \approx 10 A^3.
\ee
A crucial factor in this condition is the $e^\pm$ density. The postshock
matter emits radiation, and $e^\pm$ pairs can be produced by $\gamma$-rays 
($\gamma+\gamma\rightarrow e^-+e^+$) that have energy $h\nu>m_ec^2$ in the 
fluid frame. Suppose a fraction $\eta$ of the internal energy density 
$(A^2/2)n_bm_pc^2$ is converted into radiation and a fraction $f$ of this 
radiation is above the threshold for pair production. A maximum pair density 
is evaluated assuming that all photons emitted above the threshold $m_ec^2$ 
are converted into pairs,
\be
\label{eq:pairs}
 \frac{n_e}{n_b}\approx f\eta\frac{A^2}{2}\frac{m_p}{m_e}.
\ee
From equations~(\ref{eq:condit}) and (\ref{eq:pairs}) we conclude that
the Coulomb losses of $\alpha$-particles on $e^\pm$ can prevent
the efficient $\alpha$-$\alpha$ spallation if $f\eta>\kappa=3\times
10^{-3}AX_\alpha$. In this case, only a fraction
$\tauCoul/\taulife\approx(\kappa/f\eta)$ of the $\alpha$-particles are
destroyed. The released neutrons can then continue the spallation process
via $n$-$\alpha$ collisions (\S~4.2.1).

Equation~(\ref{eq:pairs}) assumes an optical depth for $\gamma$-$\gamma$ 
interactions $\tgg>1$. We now check this assumption. The optical depth can 
be estimated as $\tgg\approx 0.1(w_1/m_ec^2)\sT R/\Gsat$ where 
$w_1=f\eta(A^2/2)n_bm_pc^2$. This yields
\be
  \tgg\approx 0.1f\eta\frac{A^2}{2}\frac{m_p}{m_e}n_b\sT\frac{R}{\Gsat}.
\ee
It is easy to see that $\tgg>1$ at all radii where efficient 
$\alpha$-$\alpha$ spallation can take place [i.e. where 
$\taulife<(R/c\Gsat)$] if $f\eta>5\times 10^{-4}$.


\section{Conclusions}

1. Reactions of $e^\pm$ capture on nucleons, $e^-+p\rightarrow n+\nu$ and
$e^++n\rightarrow p+\bar{\nu}$, operate in GRB central engines and set
an equilibrium proton fraction $Y_e=n_p/(n_n+n_p)$. If the engine is an
accretion disk around a black hole of mass $M$, the equilibrium $Y_e$ is 
established at accretion rates 
$\dM>\dMeq\approx 10^{31}(\alpha/0.1)^{9/5}(M/M_\odot)^{6/5}$~g/s
(eq.~\ref{eq:dMeq}) where $\alpha=0.01-0.1$ is a viscosity parameter of 
the disk.

2. Of great importance for the explosion dynamics is whether $Y_e<0.5$, 
i.e., there is an excess of neutrons. A general analysis shows that
a neutrino-transparent matter has equilibrium $Y_e<0.5$ if $\mu>Q/2$,
and a similar condition for a $\nu$-opaque matter reads $\mu>Q$,
where $\mu$ is the electron chemical potential in units of $m_ec^2$ and
$Q=(m_n-m_p)/m_e=2.53$. This condition is satisfied below a critical
``neutronization'' temperature $T_n$ (eqs.~[\ref{eq:Tn_tr}] and
[\ref{eq:Tn_op}]). We find $T<T_n$ for plausible central engines of GRBs.
In particular, accretion disks have a neutron excess at
$\dM>\dM_n\approx 10^{31}(M/M_\odot)^2(\alpha/0.1)$~g/s (eq.~\ref{eq:dMn}).

3. Fireballs produced by neutron-rich engines should be also neutron rich.
A major threat for neutrons in the escaping fireball is the neutrino flux 
from the central engine as absorption of a neutrino converts the neutron
into a proton. This process is slower than the fireball expansion if the 
neutrino luminosity is below $10^{53}$~erg/s. A neutrino luminosity above
$10^{53}$~erg/s would require a very powerful and $\nu$-opaque central 
engine, which is possible. In this case, however, absorption of $\bar{\nu}$ 
by the fireball protons takes place as well as absorption of $\nu$ by the 
neutrons.  The balance between $\nu$ and $\bar{\nu}$ absorptions establishes 
a new equilibrium $Y_e$ in the fireball, which depends on the emitted 
spectra of $\nu$ and $\bar{\nu}$ and is likely below 0.5 (\S~2.3).

4. As the fireball expands and cools, the ejected free nucleons tend to
recombine into $\alpha$-particles. This process competes, however, with 
rapid expansion and can freeze out. For this reason, nucleosynthesis is 
suppressed in fireballs with a high photon-to-baryon ratio 
$\phi=n_\gamma/n_b$ (or, equivalently, high entropy per baryon 
$s/k=3.6\phi$). We find that, in radial fireballs, more than half of 
nucleons can recombine only if $\phi<3\times 10^4(r_0/3\times 10^6)^{0.85}$ 
where $r_0$ [cm] is the size of the central engine. In fireballs with 
parabolic collimation, the efficient recombination requires 
$\phi<6\times 10^5(r_0/3\times 10^6)^{0.9}$. The typical GRB parameters 
$\phi\sim 10^5$ and $r_0\sim 3\times 10^6$~cm are just marginal for 
nucleosynthesis.

5. Even in the case of efficient nucleon recombination, there are still
leftover neutrons because of the neutron excess ($Y_e<0.5$). The 
minimum mass fraction of leftover neutrons is $X_n=1-2Y_e$.

6. The nucleosynthesis also produces deuterium, which is next abundant 
element after the free nucleons and $\alpha$-particles. Its typical mass 
fraction is a few per cent. The abundances of tritium, $^3$He, and all 
elements with mass number greater than 4 are negligible.

7. Synthesized $\alpha$-particles can be spalled later on, and then the
populations of free neutrons and protons are increased. There are at least 
two possible mechanisms of spallation. (1) Energetic $n$-$\alpha$ collisions 
with a relative velocity $\tbet>\tbsp\approx 0.1$ take place before the end 
of the fireball acceleration if $\phi>4\times 10^4(\Lth/L)^{3/4}$. This 
mechanism works for fireballs with Lorentz factors 
$\Gsat>160(kT_0/1{\rm ~MeV})$ (eq.~\ref{eq:nal}) where
$T_0$ is an initial temperature (eq.~\ref{eq:T0}).
(2) Energetic $n$-$\alpha$ and $\alpha$-$\alpha$ collisions occur when the
fireball is reheated by internal shocks. This mechanism can be efficient at
modest $\Gsat<300(L_\Omega/10^{52})^{1/5}$ (eq.~\ref{eq:drift}) where
$L_\Omega$ [erg/s] is the fireball kinetic luminosity per unit solid angle.

The presence of a neutron component in GRB fireballs has quite spectacular
implications. Beside making the fireball an interesting source of multi-GeV
neutrinos (Derishev et al. 1999, Bahcall \& M\'esz\'aros 2000, M\'esz\'aros
\& Rees 2000) the neutrons survive and play a dramatic role for the
explosion development at large radii $R\sim 10^{16}-10^{17}$~cm. When the 
fireball begins to decelerate as a result of the interaction with an external 
medium, neutrons continue to coast with a high Lorentz factor $\Gamma_n$ and 
form a leading front. One can easily show that this front has kinetic energy 
much larger (by a factor of $X_n\Gamma_n\Gsat$) than the rest-mass of the 
ambient medium. It leaves behind a relativistic trail loaded with the 
products of the neutron decay until the neutron front decays completely, 
which happens at $R\approx 10^{17}$~cm 
(about $\log X_n\Gamma_n\Gsat\approx 10$ times the mean decay radius).
An external shock wave --- the customary source of GRB afterglows --- has 
to form in the neutron trail rather than in a normal ambient medium. The 
mechanism of the fireball deceleration in this situation is elaborated in 
Paper~2.

The neutron features of GRB explosions appear inevitably in the standard 
fireball scenario. They would be absent, however, if the GRBs are produced 
by magnetized winds with extremely low baryon loading, where Poynting flux 
carries much more energy than matter (e.g. Usov 1994, Lyutikov \& Blandford 
2002). A detection or non-detection of neutron effects will constrain the 
level of baryon loading in GRBs.

At the final stages of the preparation of this manuscript, a paper by Pruet,
Woosley, \& Hoffman (2002) appeared. They evaluate $Y_e$ for numerical
accretion models of Popham et al. (1999). The results are consistent with 
our analysis in \S~2 (note a slightly different definition of the viscosity 
parameter $\alpha_{\rm our}=[3/2]\alpha_{\rm Popham}$).

\acknowledgments
I thank Raymond Sawyer for discussions of the neutronization process
and Matthias Liebendoerfer for an introduction to the Lattimer-Swesty
Equation of State. I acknowledge the hospitality of the ITP at Santa Barbara,
where part of this work was done. This research was supported by NSERC, and
in part by the NSF grant PHY99-07949 and RFBR grant 00-02-16135.



\begin{references}

\reference{}
Bahcall, J. N., \& M\'esz\'aros, P. 2000, Phys. Rev. Lett., 85, 1362

\reference{}
Balbus, S. A., \& Hawley, J. F. 1998, Rev. Mod. Phys., 70, 1

\reference{}
Beloborodov, A. M. 2000, ApJ, 539, L25

\reference{}
Beloborodov, A. M. 2002, ApJL, submitted, astro-ph/0209228

\reference{}
Bulik, T., Sikora, M., \& Moderski, R. 2002, astro-ph/0209339

\reference{}
Derishev, E. V., Kocharovsky, V. V., Kocharovsky, Vl. V. 1999, ApJ, 521, 640

\reference{}
Di Matteo, T., Perna, R., \& Narayan, R. 2002, astro-ph/0207319

\reference{}
Esmailzadeh, R., Starkman, G. D., \& Dimopoulos, S. 1991, ApJ, 378, 504

\reference{}
Fuller, G. M., Pruet, J., \& Abazajian, K. 2000, Phys. Rev. Lett.,
85, 2673

\reference{}
Ginzburg, V. L., \& Syrovatskii, S. I. 1964, The Origin of Cosmic Rays,
Oxford: Pergamon

\reference{}
Kohri, K., \& Mineshige, S. 2002, ApJ, 577, 311

\reference{}
Landau, L. D., \& Lifshitz, E. M. 1980,
Statistical Physics, Pergamon Press, Oxford

\reference{}
Lattimer, J. M., \& Swesty, F. D.  1991, Nucl. Phys. A, 535, 331

\reference{}
Lemoine, M. 2002, A\&A, 390, L31

\reference{}
Lyutikov, M., \& Blandford, R. D. 2002, astro-ph/0210671

\reference{}
MacFadyen, A. I., \& Woosley, S. E. 1999, ApJ, 524, 262

\reference{}
M{\'e}sz{\'a}ros, P. 2002, ARA\&A, 40, 171

\reference{}
M\'esz\'aros, P., \& Rees, M. J. 2000, ApJ, 541, L5

\reference{}
M\'esz\'aros, P., \& Rees, M. J. 2001, ApJ, 556, L37

\reference{}
Meyer, B. S. 1994, ARA\&A, 32, 153

\reference{}
Narayan, R., Piran, T., \& Kumar, P. 2001, ApJ, 557, 949


\reference{}
Popham, R., Woosley, S. E., \& Fryer, C., 1999, 518, 356



\reference{}
Pruet, J., Guiles, S., \& Fuller, G. M. 2002, astro-ph/0205056

\reference{}
Pruet, J., Fuller, G. M., \& Cardall, C. Y. 2001, ApJ, 561, 957

\reference{}
Pruet, J., \& Dalal, N. 2002, ApJ, 573, 770

\reference{}
Pruet, J., Woosley, S. E., \& Hoffman, R. D. 2002, astro-ph/0209412

\reference{}
Qian, Y.-Z., Fuller, G. M., Mathews, G. J., Mayle, R. W., Wilson, J. R.,
\& Woosley, S. E. 1993, Phys. Rev. Lett., 71, 1965

\reference{}
Rees, M. J., \& M\'esz\'aros, P. 1994, ApJ, 430, L93

\reference{}
Ruffert, M., Janka, H.-T., Takahashi, K., \& Schaefer, G. 1997, A\&A, 319, 
122

\reference{}
Smith, M. S., Kawano, L. H., \& Malaney, R. A. 1993, ApJ, 85, 219

\reference{}
Shapiro, S. L., \& Teukolsky, S. L. 1983, Black Holes, White Dwarfs,
and Neutron Stars, (New York: Wiley-Interscience)

\reference{}
Usov, V. V. 1994, MNRAS, 267, 1035

\reference{}
Wagoner, R. V., Fowler, W. A., \& Hoyle, F. 1967, ApJ, 148, 3


\end{references}
\end{document}